\documentclass[twocolumn,tighten]{aastex631}
\usepackage{graphicx}
\usepackage{amsmath}
\usepackage{scalerel}
\usepackage{apjfonts}
\usepackage{comment}
\usepackage{array,multirow,color,tablefootnote}
\usepackage{xcolor}

\begin{document}

\title{JWST near-infrared spectroscopy of the Lucy Jupiter Trojan flyby targets: Evidence for OH absorption, aliphatic organics, and CO$_{2}$}


\author[0000-0001-9665-8429]{Ian~Wong}
\affiliation{NASA Goddard Space Flight Center, Greenbelt, MD 20771, USA; \url{ian.wong@nasa.gov}}
\affiliation{Department of Physics, American University, Washington, DC 20016, USA}

\author[0000-0002-8255-0545]{Michael~E.~Brown}
\affiliation{Division of Geological and Planetary Sciences, California Institute of Technology, Pasadena, CA 91125, USA}

\author[0000-0001-9265-9475]{Joshua~P.~Emery}
\affiliation{Department of Astronomy and Planetary Science, Northern Arizona University, Flagstaff, AZ 86011, USA}

\author[0000-0002-9995-7341]{Richard~P.~Binzel}
\affiliation{Department of Earth, Atmospheric, and Planetary Sciences, Massachusetts Institute of Technology, Cambridge, MA 02139, USA}

\author[0000-0002-8296-6540]{William~M.~Grundy}
\affiliation{Lowell Observatory, Flagstaff, AZ 86001, USA}


\author[0000-0003-2548-3291]{Simone~Marchi}
\affiliation{Southwest Research Institute, Boulder, CO 80302, USA}

\author[0000-0003-3402-1339]{Audrey~C.~Martin}
\affiliation{Department of Physics, University of Central Florida, Orlando, FL 32816, USA}

\author[0000-0002-6013-9384]{Keith~S.~Noll}
\affiliation{NASA Goddard Space Flight Center, Greenbelt, MD 20771, USA}


\author[0000-0002-9413-8785]{Jessica~M.~Sunshine}
\affiliation{Department of Astronomy and Department of Geology, University of Maryland, College Park, MD 20742, USA}

\begin{abstract}
We present observations obtained with the Near Infrared Spectrograph on JWST of the five Jupiter Trojans that will be visited by the Lucy spacecraft --- the Patroclus--Menoetius binary, Eurybates, Orus, Leucus, and Polymele. The measured 1.7--5.3~$\mu$m reflectance spectra, which provide increased wavelength coverage, spatial resolution, and signal-to-noise ratio over previous ground-based spectroscopy, reveal several distinct absorption features. We detect a broad OH band centered at 3~$\mu$m that is most prominent on the less-red objects Eurybates, Patroclus--Menoetius, and Polymele. An additional absorption feature at 3.3--3.6~$\mu$m, indicative of aliphatic organics, is systematically deeper on the red objects Orus and Leucus. The collisional fragment Eurybates is unique in displaying an absorption band at 4.25~$\mu$m that we attribute to bound or trapped CO$_2$. Comparisons with other solar system small bodies reveal broad similarities in the 2.7--3.6~$\mu$m bands with analogous features on Centaurs, Kuiper belt objects (KBOs), and the active asteroid 238P. In the context of recent solar system evolution models, which posit that the Trojans initially formed in the outer solar system, the significant attenuation of the 2.7--3.6~$\mu$m absorption features on Trojans relative to KBOs may be the result of secondary thermal processing of the Trojans' surfaces at the higher temperatures of the Jupiter region. The CO$_2$ band manifested on the surface of Eurybates suggests that CO$_2$ may be a major constituent in the bulk composition of Trojans, but resides in the subsurface or deeper interior and is largely obscured by refractory material that formed from the thermophysical processes that were activated during their inward migration.
\end{abstract}
\section{Introduction}
\label{sec:intro}

\vspace*{-\baselineskip}
\begin{deluxetable}{lcccc}[t!]
\tablewidth{0pc}
\setlength{\tabcolsep}{5pt}
\tabletypesize{\normalsize}
\renewcommand{\arraystretch}{1.1}
\tablecaption{
    Target Information
    \label{tab:targ}
}

\tablehead{ \colhead{Target} & \multicolumn{1}{c}{$D$ (km)\tablenotemark{\scriptsize a}} & \multicolumn{1}{c}{$p_{v}$ \tablenotemark{\scriptsize a}} & \multicolumn{1}{c}{Swarm}  & \multicolumn{1}{c}{Color}
}
\startdata
617~Patroclus\tablenotemark{\scriptsize b} & 146 & 0.045 & L5 & less-red \\
3548~Eurybates\tablenotemark{\scriptsize c} & 69 & 0.044 & L4 & less-red \\
21900~Orus & 60 & 0.040 & L4 & red \\
11351~Leucus & 44 & 0.037 & L4 & red \\
15094~Polymele & 22 & 0.074 & L4 & less-red \\
\enddata
\textbf{Notes.}
\vspace{-0.15cm}\tablenotetext{\textrm{a}}{Diameter $D$ and visible geometric albedo $p_{v}$ measurements are taken from the literature: \citet{mueller2010} for Patroclus, \citet{buie2018} for Polymele, \citet{buie2021} for Leucus, and \citet{mottola2023} for Eurybates and Orus.}
\vspace{-0.15cm}\tablenotetext{\textrm{b}}{In this paper, Patroclus refers to the combined Patroclus--Menoetius binary. The listed diameter is for the effective projected-area-equivalent sphere used for thermal modeling.}
\vspace{-0.15cm}\tablenotetext{\textrm{c}}{Eurybates is a collisional fragment with a significantly more neutral visible color than the other less-red objects.}
\vspace{-0.8cm}
\end{deluxetable}

\vspace*{-\baselineskip}
\vspace*{-\baselineskip}
\begin{deluxetable*}{lccccccc}[t!]
\tablewidth{0pc}
\setlength{\tabcolsep}{8pt}
\tabletypesize{\normalsize}
\renewcommand{\arraystretch}{1.1}
\tablecaption{
    JWST/NIRSpec Observation Details
    \label{tab:obs}
}

\tablehead{ \colhead{Target} & \multicolumn{1}{c}{UT Date} & \multicolumn{1}{c}{$r$ (au)\tablenotemark{\scriptsize a}} & \multicolumn{1}{c}{$R$ (au)\tablenotemark{\scriptsize a}}& \multicolumn{1}{c}{$\alpha$ (deg)\tablenotemark{\scriptsize a}} & \multicolumn{1}{c}{$V$ (mag)\tablenotemark{\scriptsize a}} & \multicolumn{1}{c}{G235M $t_{\mathrm{exp}}$ (s)\tablenotemark{\scriptsize b}}  & \multicolumn{1}{c}{G395M $t_{\mathrm{exp}}$ (s)\tablenotemark{\scriptsize b}}
}
\startdata
617~Patroclus & 2022 September 7 & 4.86 & 4.51 & 11.7 & 15.7 & 146 & 350 \\
3548~Eurybates & 2022 November 9 & 4.76 & 4.14 & 10.1 & 17.0 & 408 & 1342\\
21900~Orus & 2022 September 5 & 4.94 & 4.84 & 11.9 & 17.7 & 379 & 1167 \\
11351~Leucus & 2022 November 9 & 5.02 & 4.37 & 9.3 & 18.2 & 700 & 2889 \\
15094~Polymele & 2022 November 21 & 4.80 & 4.08 & 9.1 & 18.8 & 2101 & 3326\\
\enddata
\textbf{Notes.}
\vspace{-0.15cm}\tablenotetext{\textrm{a}}{Heliocentric distance $r$, distance from JWST $R$, phase angle $\alpha$, and $V$-band apparent magnitude of the target at the time of the JWST observation.}
\vspace{-0.15cm}\tablenotetext{\textrm{b}}{Total exposure times across both dithers in the two grating settings.}
\vspace{-0.5cm}
\end{deluxetable*}

In 2021 October,  NASA's Lucy spacecraft set out on its decade-long mission to perform the first ever flyby exploration of the Jupiter Trojans. The Trojans are confined to two vast swarms along Jupiter's orbit, centered on the L4 and L5 Lagrangian points that lead and trail the gas giant by roughly one-sixth of its orbit. These enigmatic small bodies are situated on the edge between the inner and outer solar system and hold important clues to the evolution of the entire planetary system. In the classical picture of solar system evolution, the Trojans formed either in situ or on the outer edge of the nearby main asteroid belt and should therefore bear a genetic connection to the inner solar system \citep{marzari1998,marzari2003}. In contrast, more recent dynamical instability models posit a drastically different scenario in which the Trojans originated in the primordial disk of icy planetesimals located beyond the initial orbits of the ice giants and were later scattered inward and captured into resonance during an epoch of chaotic restructuring of the giant planets' orbits \citep[e.g.,][]{gomes2005,morbidelli2005,tsiganis2005,levison2008}. Within this dynamical framework, the Trojans formed alongside the outer solar system small bodies, including the present day Kuiper belt objects (KBOs). Direct comparisons of the Trojans' observable properties with those of other small body populations promise to place the formation and evolution of Trojans definitively within our broader understanding of the early dynamical history of the solar system.

Past ground- and space-based observations showed that the reflectance spectra of Jupiter Trojans appear featureless at wavelengths below 2.5~$\mu$m \citep[e.g.,][]{dotto2006,fornasier2007,emery2011,sharkey2019,wong2019uv}, while spectroscopy at longer wavelengths revealed absorption around 3.0~$\mu$m that may be indicative of fine-grained water ice or an N--H stretch feature, as well as weaker spectral features in the 3.3--3.6~$\mu$m region that are usually associated with organics \citep{brown2016}. Photometric surveys uncovered a color bimodality that divides the Trojans into the so-called ``less-red'' and ``red'' subpopulations. The typical spectra of these two subpopulations have different 
spectral slopes at visible wavelengths that roughly correspond to the P- and D-type asteroids, respectively \citep{szabo2007,roig2008,wong2014,wong2015}. Lucy will rendezvous with five Trojan targets that sample the full diversity of surface types \citep{levison2021}. The 617~Patroclus--Menoetius binary and 15094~Polymele have colors consistent with the less-red group, while 21900~Orus and 11351~Leucus represent the red subpopulation. 3548~Eurybates is a member of a collisional family \citep[e.g.,][]{broz2011} and has a significantly more neutral visible color that is similar to that of C-type asteroids \citep{fornasier2007}. Table~\ref{tab:targ} lists the targets and their properties, in order of decreasing size.

Lucy is equipped with a low spectral resolution infrared imaging spectrometer that will obtain full-disk spectra of each Trojan target from 0.9 to 3.8~$\mu$m at a spectral resolving power of roughly 250 \citep{olkin2021}. At close approach, Lucy will carry out spatially resolved scans of portions of the surface at a spatial resolution of a few kilometers, as well as multicolor imaging at higher spatial resolution. These datasets will provide a unique opportunity to explore the detailed chemical composition and surface geology of these Trojans at fine spatial scales. In preparation for the encounters in 2027--2033, a multifaceted campaign of ground- and space-based observations is being undertaken to maximize our understanding of the Trojan targets prior to the flybys and compile complementary datasets that will be indispensable for the interpretation of the Lucy observations. 

As part of this effort, near-infrared spectroscopic observations of the five Lucy Trojan targets were obtained at a spectral resolving power~$\sim$~1000 with the Near-Infrared Spectrograph (NIRSpec) instrument \citep{jakobsen2022,boker2023} on JWST. These high-quality 1.7--5.3~$\mu$m spectra are ideally suited to probe the wavelength regions where previous detections of spectral features have been reported, while also significantly expanding the wavelength coverage beyond 4~$\mu$m into a region that is out of reach of both ground-based observatories and Lucy's instruments. In this manner, the JWST observations will help paint a more complete picture of the Lucy flyby targets. By placing these JWST spectra within the geological context that Lucy will provide, we will obtain a powerful tool for understanding the surface composition, geology, and evolutionary history of these objects. More broadly, the insights gained from a joint analysis of the JWST and Lucy flyby data will permit significantly more sophisticated interpretations of the myriad future JWST observations of asteroids, irregular satellites, and KBOs, most of which will never benefit from spacecraft visits. 


\section{JWST Observations}
\label{sec:obs}

Near-infrared observations of the five Trojan targets were obtained using the NIRSpec integral field unit (IFU) as part of JWST Cycle 1 General Observers Program \#2574 (PI: M. Brown). All five visits utilized the same observing strategy: after slewing and instrument setup, a pair of dithered exposures was collected with the G395M grating, followed immediately by a pair of shorter-duration exposures with the G235M grating. These two gratings provide continuous, overlapping wavelength coverage from 1.7 to 5.3~$\mu$m at a spectral resolving power of around 1000. The observations employed the IRS$^2$ readout method, which reduces the level of correlated noise caused by electronic drifts during detector readout (so-called $1/f$ noise; \citealt{moseley2010,rauscher2012}). Details of the observations are summarized in Table~\ref{tab:obs}.

The uncalibrated data products were processed through a custom spectral extraction pipeline. The conversion from the raw 2D detector readouts to fully calibrated 3D spectral data cubes was handled by v1.11.3 of the official JWST calibration pipeline \texttt{jwst} \citep{bushouse2022}. The calibration reference files required for bias subtraction, flat-fielding, wavelength calibration, flux calibration, and other relevant data processing routines were automatically downloaded from context \texttt{jwst\_1100.pmap} of the JWST Calibration Reference Data System. Several adjustments were made to the default data processing workflow. After the stacks of nondestructive detector readouts (\texttt{uncal} files) were converted to 2D countrate images (\texttt{rate} files), we manually corrected for residual low-level read noise by subtracting the 200 pixel wide moving median of the off-sky pixels from each detector column. This process removed the visible column striping that typically persists on NIRSpec full detector images even when the IRS$^2$ readout method is selected. After the spectral data cubes were constructed for each dithered exposure, we did not proceed with the final stage of the \texttt{jwst} pipeline, which aligns the moving target between dither positions and generates a single combined data cube for each NIRSpec grating setting. Instead, we extracted the spectrum from each dithered exposure separately, which allowed us to inspect the individual spectra to examine minor spectral features and distinguish them from correlated noise or other low-level instrumental artifacts.

\begin{figure}
    \centering
    \includegraphics[width=0.8\columnwidth]{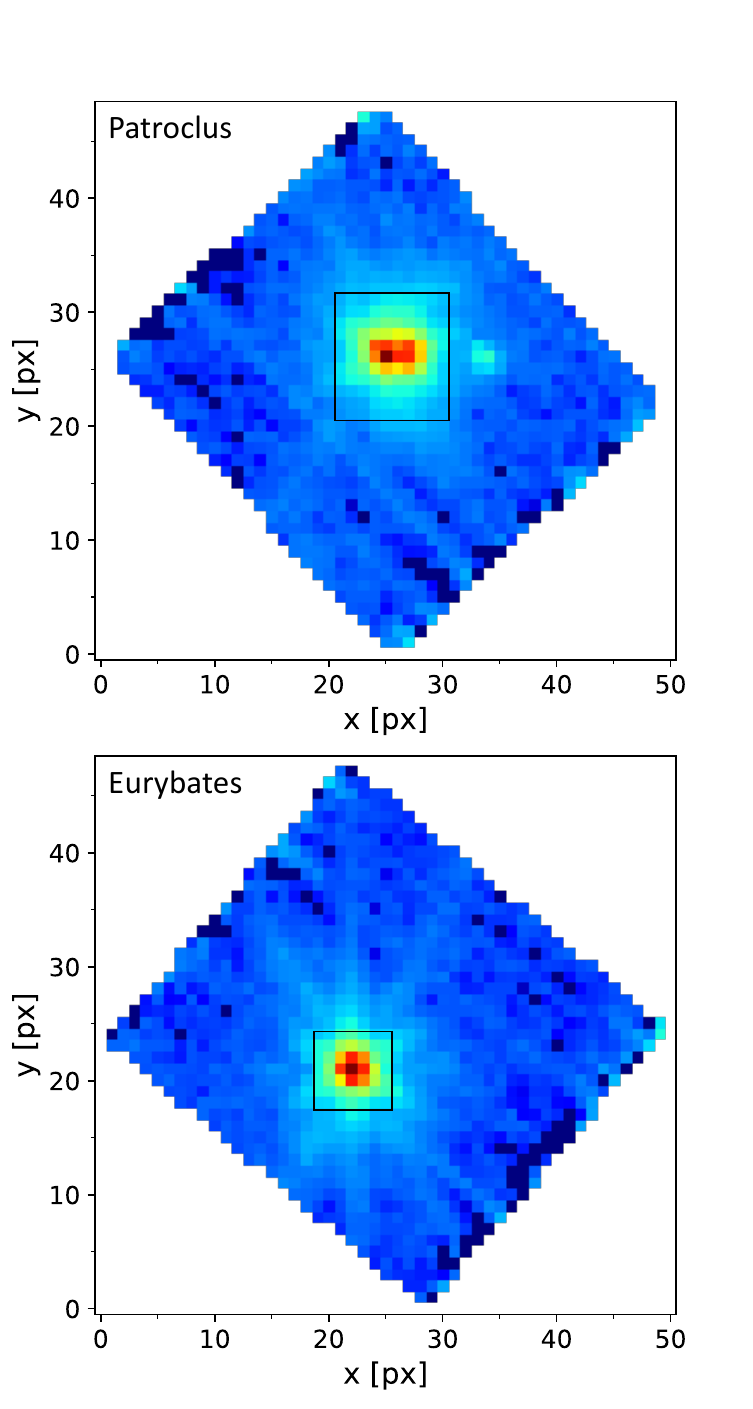}
    \caption{Examples of median-averaged IFU slices obtained with the G235M grating setting for Patroclus and Eurybates. The slices are projected onto the sky plane, with the $x$- and $y$-coordinates corresponding to R.A. and decl., respectively. A logarithmic stretch has been applied to the flux scaling in order to accentuate the detailed structure of the target PSFs at wide separations. The PSF of the Patroclus binary shows an extended shape, with the two binary components blended together. The black boxes denote the extraction regions used in producing the spectra. The faint background source located roughly 7~pixels to the right of Patroclus lies in a region of the field of view that was masked during our spectral extraction.}
    \label{fig:ifu}
\end{figure}

\begin{figure*}
    \centering
    \includegraphics[width=\textwidth]{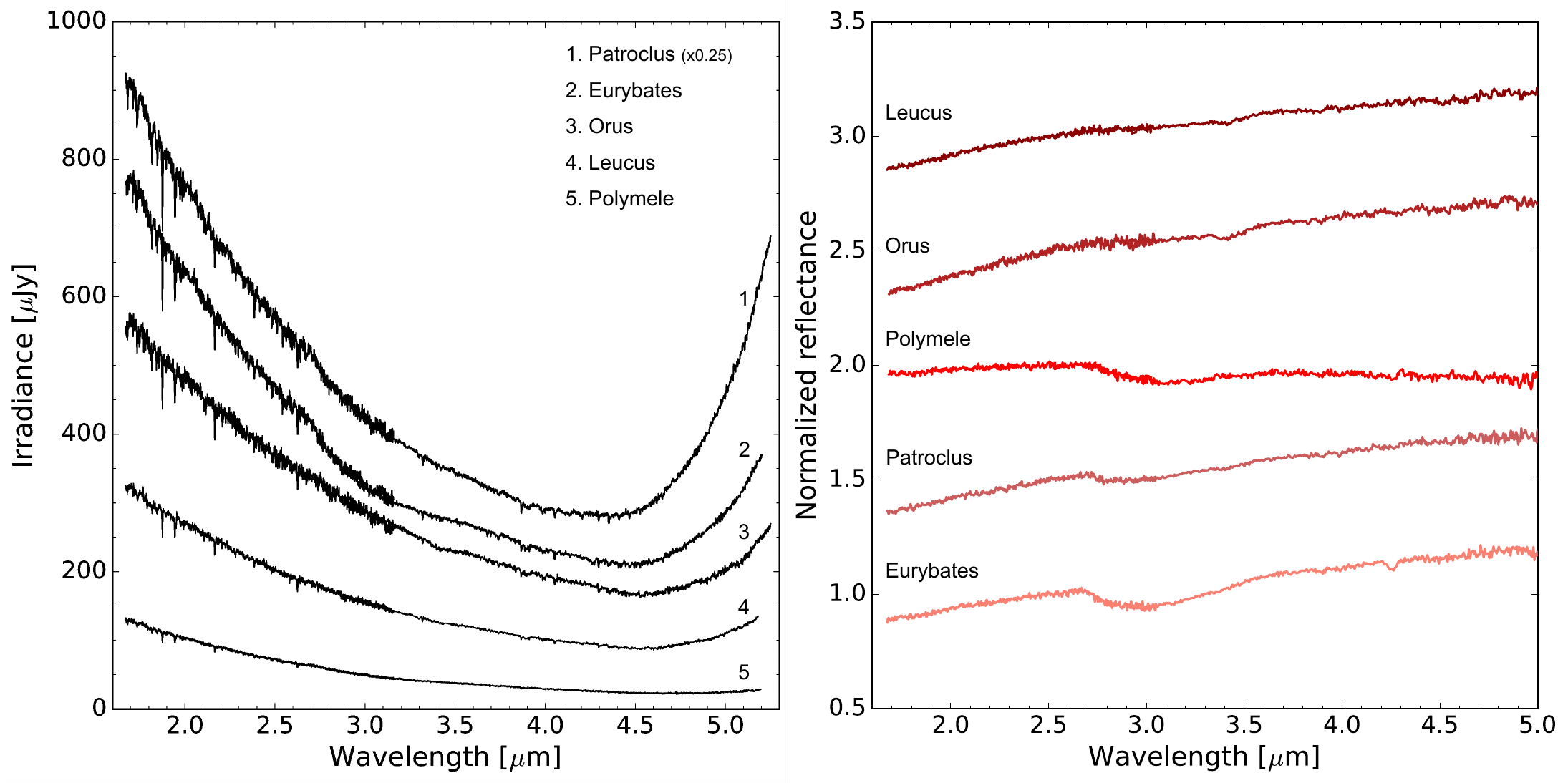}
    \caption{Left: extracted irradiance spectra of the five Trojan targets. The spectrum of Patroclus has been scaled by 0.25. The large-scale structure of each spectrum is dominated by reflected sunlight below 3~$\mu$m and the target's thermal emission beyond 4.5~$\mu$m. Right: reflectance spectra of the five Trojan targets, with the thermal tail removed and normalized to unity at 2.5~$\mu$m. The JWST Trojan reflectance spectra shown here and in all subsequent figures are binned by a factor of 3. The spectra are offset by 0.5 for clarity and arranged in order of increasing spectral slope as measured at visible wavelengths. The calculated uncertainties are generally comparable to the local scatter in the spectra.}
    \label{fig:spectra}
\end{figure*}


The output IFU data cubes (\texttt{s3d} files) contain a stack of 2D wavelength slices with a $3'' \times 3''$ field of view and a spatial pixel scale of $0\overset{''}{.}1$. Figure~\ref{fig:ifu} shows the median-averaged slices of Patroclus--Menoetius and Eurybates from the first dithered exposure with the G235M grating; the slices have been resampled to the sky-projected coordinate frame by the JWST pipeline. The point-spread function (PSF) of the Patroclus--Menoetius binary has a discernibly extended shape. According to the binary orbit ephemerides in \citet{grundy2018}, Menoetius was situated roughly $0\overset{''}{.}25$ (i.e., 2.5~pixels) to the west of Patroclus at the time of observation (toward the right in Figure~\ref{fig:ifu}). Meanwhile, the additional source apparent on the image roughly 7 pixels to the right of the binary is a background object. We experimented with using the \texttt{webbpsf} package \citep{perrin2014} to simulate the combined PSF of the binary, but found that the generated PSFs were not an adequate match to the observed PSFs, resulting in significant residuals in the modeling and unreliable flux extraction for the two binary components. Instead, we carried out spectral extraction on the blended binary as a single source. Previous differential photometry of Patroclus and Menoetius during mutual events did not reveal any significant systematic differences in color between the binary components \citep{wong2019patroclus}, suggesting large-scale similarity in surface composition. Hereafter, we refer to the binary as Patroclus for the sake of brevity. The remaining four targets show radially symmetric PSFs that are consistent with point sources. In the case of Orus, the target was blended with a background star during one of the two G235M exposures, and the corresponding data cube was not included in our analysis.

Spectral extraction was carried out using an empirical PSF fitting technique that has been applied to other NIRSpec IFU observations of minor bodies \citep[see, for example,][for more detailed descriptions of the methodology]{brown2023,emery2023,grundy2023}. The source PSF in each wavelength slice $j$ was modeled from the median frame that was constructed using a sliding window spanning $j \pm 10$. The width of this window was chosen to be sufficiently wide to average out interpixel jitter and uncorrected bad pixels, while simultaneously being narrow enough to reflect the local shape of the PSF accurately, which exhibits measurable shape oscillations across the wavelength range due to undersampling of the PSF at the NIRSpec IFU pixel scale, drifts in the centroid of the dispersed spectral trace across detector rows, and corresponding resampling artifacts in the cube building process. The background region was defined as all pixels outside of a $21 \times 21$ pixel box centered on the target centroid position in the IFU field of view, and the median background flux was subtracted from the median frame to generate the PSF template. This template was set to zero in the background region and renormalized to a sum of unity, before being fit to the wavelength slice $j$ with a flux scaling factor and an additive background level.

We optimized the size of the extraction region within which the PSF template was defined to produce the cleanest spectrum. We found that including the extended diffraction wings of the source PSF (see Figure~\ref{fig:ifu}) contributes additional scatter to the resultant spectra, so we selected extraction regions that primarily sampled the peak of the source PSF, where the signal-to-noise ratio is the highest and the PSF templates can be reliably modeled. Pixels outside of the extraction region but inside the background region were masked during the fitting process. For Patroclus, we used a square extraction region of side length 11 pixels in order to encompass the PSFs of both binary companions. For Eurybates, Orus, and Leucus, a $7 \times 7$ pixel box was chosen, while for the faintest target Polymele, a smaller $5 \times 5$ pixel yielded reduced correlated noise in the extracted spectrum when compared to larger apertures. The pairs of dithered exposure spectra were passed through a 20 pixel wide moving median filter to remove $3\sigma$ outliers before being averaged together to produce the final combined irradiance spectra. The irradiance spectra of the five Trojan targets are plotted in the left panel of Figure~\ref{fig:spectra}. At short wavelengths, the irradiance closely mirrors the shape of the incident sunlight spectrum, while at $\lambda > 4.5$~$\mu$m, the thermal emission from the targets' surfaces dominates. At the spectral resolution of the G235M and G395M gratings, solar lines are also visible throughout the wavelength range.

To derive the reflectance spectra, we divided the targets' irradiance spectra by the spectrum of the G2V-type solar analog star P330E, which was observed as part of Program \#1538 (PI: K. Gordon). Each target spectrum was paired with a stellar spectrum that was produced with the same extraction aperture size. This process self-consistently corrected for both common-mode instrumental systematics present in the target and stellar spectra and wavelength-dependent flux biases due to the monotonically increasing size of the source PSF within our fixed extraction apertures. There are strong correlated noise artifacts at the long-wavelength end of the G235M spectra, and we removed the portions between 3.05 and 3.15~$\mu$m prior to analyzing and plotting the reflectance spectra.

The spectra of all targets show significant contributions from thermal emission at wavelengths longer than $\sim$4~$\mu$m.  We modeled and removed this thermal contribution using the Near Earth Asteroid Thermal Model (NEATM; \citealt{harris1998}), following the methods described in \citet{mcgraw2022}. When implementing NEATM, we fixed the targets' sizes and albedos to literature values (see Table~\ref{tab:obs}). We assumed a linear reflectance continuum, nominally pinned to the spectrum at 2.6~$\mu$m, and varied the beaming parameter in the model to determine the level of thermal emission that best fits the measured spectrum at longer wavelengths. This model thermal contribution was then subtracted to produce the final reflectance spectra. We also varied the slope and vertical position of the assumed reflected continuum to evaluate the range of reasonable spectral shapes. In all cases, fixing the continuum slope to that of a line between 2.6 and 3.7~$\mu$m (wavelengths that are dominated by reflected flux) minimized the resultant residuals in the modeled spectra. After removal of the thermal component, some artifacts became apparent at the longest wavelengths, manifesting as unphysical breaks in the continuum shape. These artifacts were likely caused by instrumental systematics that were not corrected by dividing the solar analog spectrum, and we trimmed the spectra beyond 5.0~$\mu$m. 

\vspace*{-\baselineskip}
\vspace*{-\baselineskip}
\begin{deluxetable*}{lccccccc}[t!]
\tablewidth{0pc}
\setlength{\tabcolsep}{4pt}
\tabletypesize{\normalsize}
\renewcommand{\arraystretch}{1.1}
\tablecaption{
    Results of the Absorption Band Analysis
    \label{tab:bands}
}
\tablehead{  & \multicolumn{3}{c}{3.0~$\mu$m OH Band} & \multicolumn{2}{c}{3.4~$\mu$m Organic Band} & \multicolumn{2}{c}{4.25~$\mu$m CO$_2$ Band} \\
Target & Center ($\mu$m) & Depth (\%) & 2.9~$\mu$m Depth (\%) & Center ($\mu$m) & Depth (\%) & Center ($\mu$m) & Depth (\%) 
}
\startdata
617~Patroclus & $3.031 \pm 0.010$ & $3.62 \pm 0.30$ & $3.51 \pm 0.34$ & $3.409 \pm 0.015$ & $1.00 \pm 0.20$ & $\dots$ & $\dots$\\
3548~Eurybates & $3.028 \pm 0.003$ & $8.83 \pm 0.24$ & $8.12 \pm 0.27$ & $3.385 \pm 0.010$ & $1.00 \pm 0.20$ & $4.258 \pm 0.002$ & $3.64 \pm 0.33$ \\
21900~Orus & $3.271 \pm 0.026$ & $1.18 \pm 0.26$ & $0.53 \pm 0.33$ & $3.428 \pm 0.008$ & $2.78 \pm 0.37$ & $\dots$ & $\dots$ \\ 
11351~Leucus & $3.223 \pm 0.034$ & $1.05 \pm 0.22$ & $0.06 \pm 0.25$ & $3.440 \pm 0.015$ & $1.87 \pm 0.40$ & $\dots$ & $\dots$\\
15094~Polymele & $3.088 \pm 0.012$ & $6.61 \pm 0.41$ & $4.32 \pm 0.44$ & $\dots$ & $\dots$ & $\dots$ & $\dots$
\enddata
\end{deluxetable*}

The wavelength grid of the IFU data cubes generated by the JWST calibration pipeline is oversampled by a factor~$\sim$~2.2 relative to the true spectral resolution of the G235M and G395M gratings.\footnote{Tabulated dispersion and resolution curves for all NIRSpec gratings can be found at \url{https://jwst-docs.stsci.edu/jwst-near-infrared-spectrograph/nirspec-instrumentation/nirspec-dispersers-and-filters}.} We used median binning to downsample the spectra by a factor of 3. This binning factor is used for all figures displaying the JWST Trojan reflectance spectra. The final binned and thermal-corrected reflectance spectra are plotted in the right panel of Figure~\ref{fig:spectra}; here, the targets are arranged in order of increasing visible spectral slope (i.e., redness). The flux uncertainties computed in our PSF template fitting procedure are comparable to the inherent scatter in the resultant spectra. The spectra obtained with the G395M grating were scaled to match the reflectance level in the G235M spectra at 3.0~$\mu$m. For most targets, the required scaling factor was small ($<0.2\%$), while for Orus a larger $\sim$2\% downward adjustment of the G395M spectrum was needed, which can be attributed to low-level systematic flux offsets between the two dither positions and the aforementioned removal of one of the two G235M dithered exposures due to background star blending.

\section{Surface Characterization}
\label{sec:char}
The JWST/NIRSpec observations represent a substantial leap in our knowledge of Trojan spectra. Previous observations of these five Trojan targets have largely been confined to wavelengths below 2.5~$\mu$m. \citet{sharkey2019} presented 0.7--2.5~$\mu$m reflectance spectra obtained with the NASA Infrared Telescope Facility (IRTF). In the overlapping wavelength range (1.7--2.5~$\mu$m), most of their spectra have continuum slopes that agree with our JWST spectra. For Polymele, the IRTF spectrum shows a somewhat steeper median slope than what we obtained from JWST; however, the former spectrum has a very low signal-to-noise ratio within the wavelength range of interest, allowing for a wide range of spectral slopes, including neutral values that are fully consistent with the continuum shape shown in Figure~\ref{fig:spectra}.

Our spectra offer improved precision relative to previously published results and expand our purview beyond 4~$\mu$m into a hitherto unexplored region of wavelength space. The reflectance spectra display several distinct spectral features across the 2.5--4.5~$\mu$m wavelength range. In this section, we characterize these features through comparisons with published results of other minor bodies, with the goal of identifying plausible chemical species on the surfaces and placing the Trojans within the broader context of inner and outer solar system small body populations.

\subsection{3~$\mu$m OH Band}
\label{subsec:3micron}
The most prominent feature in the Trojan spectra is the broad absorption band that is centered around 3~$\mu$m. This band is deepest on Eurybates and clearly discernible on the other less-red objects Patroclus and Polymele. Meanwhile, on the red objects Orus and Leucus, the presence of this band is considerably more marginal. A characteristic property of the absorption is its distinct short-wavelength edge at roughly 2.7~$\mu$m.  Previously, \citet{brown2016} detected a portion of this 3~$\mu$m band in Keck/NIRSPEC spectra of 16 Trojans, with less-red objects displaying deeper band depths than red objects. The significantly higher signal-to-noise ratio JWST/NIRSpec spectra confirm the earlier detections and robustly corroborate the trend between surface color and 3~$\mu$m band depth. 

\begin{figure}
    \centering
    \includegraphics[width=\columnwidth]{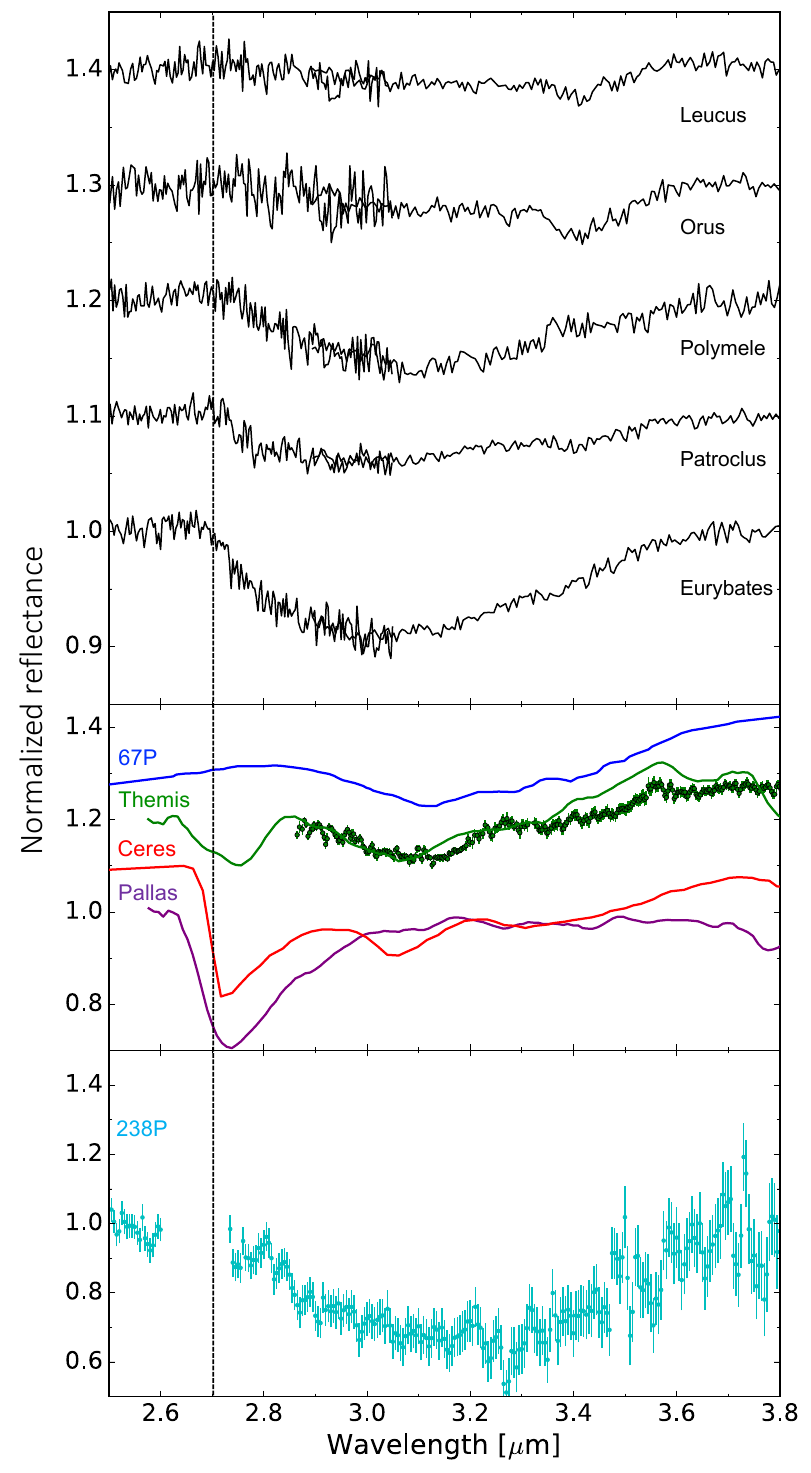}
    \caption{Top panel: binned continuum-subtracted reflectance spectra of the five Trojan targets in the 2.5--3.8~$\mu$m region, normalized to unity at 2.5~$\mu$m. The less-red objects Eurybates, Patroclus, and Polymele show a significant 3~$\mu$m spectral feature indicative of OH absorption. This absorption has a characteristic short-wavelength edge around 2.7~$\mu$m, which is marked by the vertical dashed line. Middle panel: published reflectance spectra of the main belt asteroids Pallas (purple; \citealt{usui2019}), Ceres (red; \citealt{desanctis2018}), and Themis (green curve from \citealt{usui2019}; green points from \citealt{rivkin2010}), as well as the Jupiter-family comet 67P (blue; \citealt{poch2020}). All spectra in this panel are normalized to unity at 2.6~$\mu$m. The shapes of the 2.6--3.2~$\mu$m bands in these spectra do not match the 3~$\mu$m band seen on Trojans, indicating different surface compositions. Bottom panel: the reflectance spectrum of the active outer main belt asteroid 238P, obtained with JWST/NIRSpec by \citet{kelley2023} and normalized to unity at 2.5~$\mu$m, which shows a broad 2.7--3.7~$\mu$m absorption that is similar to the features seen on the less-red Trojans. The 2.6--2.7~$\mu$m water vapor emission band from the coma has been trimmed.}
    \label{fig:3micron_1}
\end{figure}

We determined the center of this broad absorption by dividing each spectrum by a linear continuum fit across the feature, fitting a third-order polynomial to the bottom of the band, and taking the location of the minimum of the best-fit polynomial as the band center. For Orus and Leucus, we excluded the region around the 3.4~$\mu$m organic band from the polynomial fit (see Section~\ref{subsec:3.4micron}). The band depth was calculated by averaging the continuum-divided spectrum within a 0.05~$\mu$m wide window centered around the band center. This window width was selected so that $\sim$25--30 channels are averaged. This process was repeated 10,000 times, each time resampling the data points using normal distributions with widths equal to the respective uncertainties. The reported values and uncertainties are the mean and standard deviation of the resulting parameter distributions. We also computed the band depth at 2.9~$\mu$m to enable more direct comparisons with ground-based spectra of near-Earth and main belt asteroids, whose band depths are often reported at this wavelength in the literature. Table~\ref{tab:bands} shows the results of our band analysis. The continuum-subtracted Trojan spectra in this region are presented in Figure~\ref{fig:3micron_1}.

The continuous wavelength coverage of the JWST spectra enables us to perform a more detailed characterization of the 3~$\mu$m band than was previously possible. High-quality spectra of small bodies covering the 3~$\mu$m region are scant in the literature. Ground-based observations in this wavelength range are challenging due to significant telluric absorption, which almost entirely suppresses the incoming light at 2.5--2.9~$\mu$m. While a significant body of IRTF/SpeX spectra is available for main belt asteroids \citep{takir2012}, including many that show superficially similar 3~$\mu$m absorption features to the Trojans, the gap in wavelength coverage precludes definitive compositional inferences. The only available continuous reflectance spectra in this region were obtained from space telescopes or spacecraft flyby missions. In Figure~\ref{fig:3micron_1}, we plot a collection of representative spectra that show various types of absorption features that have been observed on main belt asteroids.

The spectrum of Pallas, derived from measurements made by the AKARI spacecraft \citep{usui2019}, shows a distinctive sharp 2.7~$\mu$m absorption that is diagnostic of hydrated silicates --- specifically, the O--H stretch feature in phyllosilicates \citep[e.g.,][]{lebofsky1980,jones1990,rivkin2002}. The spectra of Ceres (from globally averaged observations by the Dawn spacecraft; \citealt{desanctis2018}) and Themis (compiled from AKARI and ground-based IRTF spectra; \citealt{rivkin2010,usui2019}) display the same 2.7~$\mu$m hydrated silicate band. For Trojans generally, we can exclude hydrated silicates as a major component in the surface regolith due to the absence of a sharp absorption band centered at $\sim$2.7--2.8~$\mu$m as seen in Ceres, Palas, and Themis. 

Themis and Ceres show a second spectral feature around 3.0--3.2~$\mu$m; at even longer wavelengths, there are other features from organics and/or carbonates, which will be discussed in the following subsection. Several interpretations have been forwarded for the 3.0--3.2~$\mu$m band. In the case of Themis models incorporating very fine water frost ($<0.1$~$\mu$m) coating underlying refractory grains provide a good match (\citealt{campins2010}; \citealt{rivkin2010}; but see \citealt{mckay2017}; \citealt{orourke2020} for arguments against this interpretation). Meanwhile, N--H stretch absorption from ammoniated surface species has been hypothesized to be a primary contributor to the feature on Ceres \citep[e.g.,][]{king1992,desanctis2015}. The lack of absorption around 3.0--3.2~$\mu$m in the Trojan spectra leads us to argue against the presence of fine-grained water frost or ammoniated species as a distinct surface element on Trojans.

We have also included the spectra of two active objects: the Rosetta spectrum of the Jupiter-family comet 67P/Churyumov--Gerasimenko \citep[67P;][]{poch2020} and the recently published JWST/NIRSpec spectrum of the active outer main belt asteroid 238P/Read \citep[238P;][]{kelley2023}. 67P lacks the hydrated silicate feature at 2.7~$\mu$m while having a complex of absorption bands spanning 2.8--3.6~$\mu$m that is attributed to a combination of ammoniated salts and organics. In contrast, the spectrum of 238P (shown separately in the bottom panel of Figure~\ref{fig:3micron_1}) has a single broad 2.7--3.6~$\mu$m band that is quite comparable to the feature seen in the less-red Trojan spectra, albeit with the reflectance minimum situated at somewhat longer wavelengths. Although 238P is much smaller ($\sim$0.6~km) than the Trojans there were observed by JWST, the similarity in spectral shapes is noteworthy. 238P is more dynamically akin to typical outer main belt asteroids than to the Jupiter-family comets, indicating that it was not recently implanted into the inner solar system from the Kuiper belt \citep{haghighipour2009,hsieh2016}. Nevertheless, it is still possible that the origin of 238P lies in the outer Solar System, as dynamical instability models predict a nonnegligible contingent of primordial trans-Neptunian planetesimals among the outer main belt asteroid population \citep[e.g.,][]{vokrouhlicky2016}. Future spectroscopic surveys of more small outer main belt asteroids may uncover other objects with Trojan-like 2.7--3.6~$\mu$m absorption bands.



Given the predictions from dynamical instability models that place the origin of the Trojans in the outer solar system, we now turn to KBOs for further insight into the 3~$\mu$m feature. Prior to JWST, only a handful of KBOs had been spectroscopically studied at wavelengths longer than 3~$\mu$m. During the first year of science operations, JWST observed more than 70 KBOs, providing a treasure trove of 1--5~$\mu$m reflectance spectra spanning all dynamical classes and a wide range of sizes. For comparison with our Trojan spectra, we analyzed publicly available NIRSpec observations of a pair of KBOs that were obtained as part of General Observers Program \#2418 (PI: N. Pinilla-Alonso; \citealt{pinillaalonso2023}). Photometric surveys of KBOs have shown that objects smaller than 1000~km in diameter can be broadly divided into two groups based on their visible colors \citep[e.g.,][]{fraser2012,wong2017,fraser2023}. The two KBO targets we selected---2004~NT33 (420~km; \citealt{vilenius2014}) and 2004~TY364 (510~km; \citealt{lellouch2013})---are respectively from the so-called red and very-red subgroups within the hot classical population. 

\begin{figure}
    \centering
    \includegraphics[width=\columnwidth]{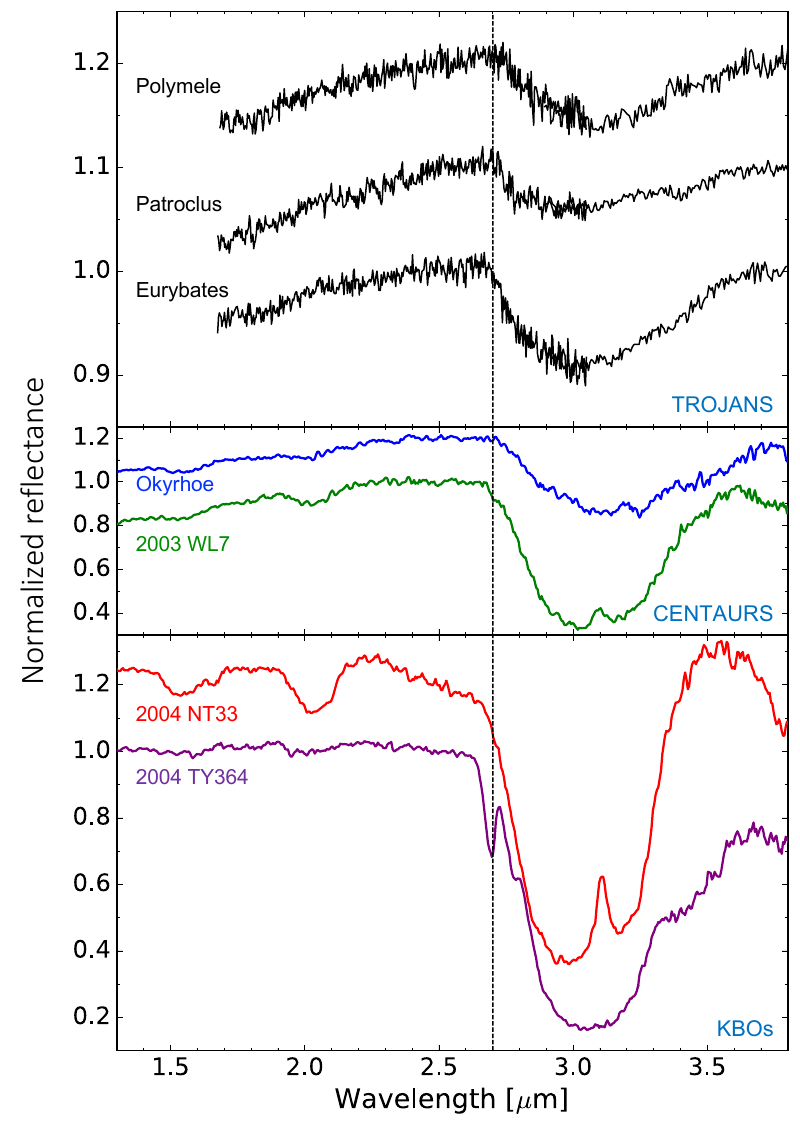}
    \caption{Top panel: binned 2.5--3.8~$\mu$m reflectance spectra of the less-red Trojans Eurybates, Patroclus, and Polymele. Middle panel: JWST reflectance spectra of the Centaurs Okyrhoe and 2003~WL7. Bottom panel: JWST reflectance spectra of the red and very-red hot classical KBOs 2004~NT33 and 2004~TY384. All spectra have been normalized to unity at 2.5~$\mu$m and offset for clarity. All of these spectra display a broadly similar 3~$\mu$m OH band shape, with a short-wavelength edge at $\sim$2.7~$\mu$m (vertical dashed line). Okyrhoe's spectrum bears the strongest resemblance to the less-red Trojan spectra, suggesting a comparable surface composition.}
    \label{fig:3micron_2}
\end{figure}

In addition, we included two Centaurs (i.e., inward-scattered KBOs that cross the giant planet region on unstable orbits) that were observed as part of Program \#2418 and published in \citet{licandro2023}: Okyrhoe and 2003~WL7 (35 and 103~km, respectively; \citealt{duffard2014}). These Centaurs have perihelion and aphelion distances of $q = 5.8$~au and $Q = 8.4$~au, and $q = 14.9$~au and $Q = 20.1$~au, respectively. All four observations were obtained with the low spectral resolution prism grating (0.6--5.3~$\mu$m; $\Delta\lambda/\lambda$~$\sim$~30--300), and we processed the observations following the same methodology that was outlined in Section~\ref{sec:obs}. The prism-mode spectrum of the G-type solar analog star SNAP-2 (Program \#1128; PI: N. Luetzgendorf; publicly available data) was used to produce the reflectance spectra, which we plot in Figure~\ref{fig:3micron_2}. 

The spectra of 2004~NT33 and 2004~TY364 differ broadly in several ways. 2004~NT33 shows strong water ice absorption bands at 1.5, 2, and 3~$\mu$m, including a distinct feature at 1.65~$\mu$m and the Fresnel reflection peak at 3.1~$\mu$m, both of which are indicative of the crystalline form of water ice (see \citealt{mastrapa2013} for a review of amorphous and crystalline water ice on solar system objects). Meanwhile, the spectrum of 2004~TY364 is considerably less water ice rich, with only marginal detections of the 1.5 and 2~$\mu$m absorption features and no signs of the Fresnel peak. It shows a deep and broad 2.7--3.3~$\mu$m absorption feature and an additional rounded spectral band at 3.3--3.6~$\mu$m that corresponds to aliphatic organics (see Section~\ref{subsec:3.4micron}). The pair of narrow features between 2.6 and 2.8~$\mu$m are overtones of the $\nu_3$ band of CO$_2$ ice \citep[e.g.,][]{quirico1997,baratta1998}, which have been detected in the JWST spectra of many KBOs \citep{depra2023,pinillaalonso2023}. Moving inward to the Centaurs, we find that the spectrum of 2003~WL7 is largely consistent with 2004~NT33, showing clear water ice absorption bands, albeit with somewhat reduced depths. In contrast, the water ice features on Okyrhoe are strongly attenuated, and the shallow 3~$\mu$m band has a more asymmetric profile that gently rises toward the continuum at 3.6~$\mu$m. 

The four KBO and Centaur spectra shown in Figure~\ref{fig:3micron_2} are broadly representative of the range of 3~$\mu$m band shapes seen on similarly sized bodies throughout the trans-Neptunian region \citep{licandro2023,pinillaalonso2023}. Placing these spectra alongside our results for the Trojans, we note the strong similarity in the general shape and width of the broad 3~$\mu$m absorption features to the observed feature on Eurybates, Patroclus, and Polymele, with consistent short-wavelength edges located at roughly 2.7~$\mu$m. This comparison suggests that the same molecular vibration --- namely, the fundamental O--H stretch mode \citep[e.g.,][]{mastrapa2009} --- is responsible for the 3~$\mu$m absorption across all of these objects. In the case of KBOs and Centaurs, the primary contributor to the 3~$\mu$m band is water ice (both amorphous and crystalline), which is known to be widely distributed across the populations based on detections of the 1.5 and 2~$\mu$m bands \citep[e.g.,][]{barkume2008,barucci2011}. 

The identity of the OH-bearing component on the Trojans' surfaces is less clear. The absence of the 2~$\mu$m water ice band precludes a significant water ice fraction across the surface, consistent with previous observations \citep{emery2003,yang2007}. This finding also aligns with the results from thermophysical modeling of Trojans, which demonstrate that water ice should be depleted down to $\sim$10~m depth on average, though the sublimation front at polar regions may be considerably shallower, up to a depth~$\sim$~10~cm \citep{guilbertlepoutre2014}. While a coating of fine-grained water frost similar to the proposed component on Themis can effectively suppress the 1.5 and 2~$\mu$m absorption features \citep{rivkin2010}, it produces an OH band at 3.1~$\mu$m that is too narrow to match the observed feature on Trojans. Instead, larger micron-sized grains of water ice are necessary to reproduce the broad rounded shape and the short-wavelength edge at 2.7~$\mu$m that are seen on Trojans, KBOs, and Centaurs, as well as the comets Tempel~1 \citep{sunshine2006}, Hartley~2 \citep{protopapa2014}, and 238P \citep{kelley2023}.

Exposure of more pristine subsurface material through impact gardening has been proposed as a possible avenue by which otherwise unexpected water ice features may emerge on small dark bodies in the middle solar system. Spatially resolved observations of the irregular Saturnian satellite Phoebe, which has a similar size to the largest Trojans and is thought to have originated from the outer solar system, have shown that while water ice absorption is manifested across the entire surface, they are most prominent within impact basins and are correlated with patches of higher optical geometric albedo \citep{clark2005,buratti2008,fraser2018}. 

Extending this argument to the Trojans, which have experienced a similar level of collisional evolution as the main belt asteroids \citep{marchi2023}, surface turnover from cratering impacts may periodically expose small localized patches of water-rich interior material amid the predominant water-ice-poor dark regolith. However, it is unclear if the present day impactor flux is sufficient to maintain pristine water ice on the surface, given the significantly higher rate of sublimation loss at the higher temperatures of the Trojan region \citep{guilbertlepoutre2014,lisse2022}. We note that isolated patches of water ice have been detected in spatially resolved spectra of the comets Tempel~1 \citep{sunshine2006} and 67P \citep{barucci2016}, which experience much higher average surface temperatures than the Trojans. In those cases, the water ice is concentrated in regions that are topographically shielded from insolation or freshly exposed due to physical processes such as mass wasting. It follows that analogous surface configurations on the Trojans may preserve water ice for extended periods. In the context of this proposed scenario, the difference between the 3~$\mu$m spectra of less-red and red Trojans would indicate that the former have intrinsically more water ice in their subsurface than the latter and suggests distinct initial surface compositions. The future Lucy flybys will provide spatially resolved optical and near-infrared photometry and spectroscopy of the Trojan targets' surfaces and allow us to assess whether or not localized regions of exposed water ice are present on these objects.

An alternative explanation for the observed 3~$\mu$m OH band on the less-red Trojans is organic refractory material formed by the irradiation of carbon-bearing ices. Laboratory experiments simulating the effect of solar processing on H$_2$O~$+$~CH$_3$OH~$+$~NH$_3$ ice mixtures \citep{urso2020} produced complex residues that display a broad, asymmetric absorption band extending from 2.7--2.8~$\mu$m to 3.6--3.8~$\mu$m, in good agreement with the absorption feature seen on Trojans. This interpretation has also been presented to explain the absorption feature observed on 238P \citep{kelley2023}. In this scenario, the OH band around 3~$\mu$m is produced through the dissociation of H$_2$O and CH$_3$OH. Additional bands that contribute to extending the overall absorption feature to longer wavelengths are due to C--H and N--H stretch modes. 

The KBO--Centaur--Trojan comparison has important implications for our understanding of solar system dynamical evolution. Numerical integrations that simulate the effect of current dynamical instability models on the primordial trans-Neptunian disk invariably show that the inward-scattered planetesimals, some of which ultimately become captured into resonance by Jupiter, pass through an intermediate stage when they lie on highly eccentric giant-planet-crossing orbits \citep[e.g.,][]{morbidelli2005,nesvorny2013,roig2015}. Centaurs therefore serve as an ideal proxy for assessing how Trojans may have evolved during their inward migration. 

The spectrum of Okyrhoe is unique among the body of published JWST spectra of KBOs, with its extremely muted ($<20\%$) 3~$\mu$m OH band and the characteristic asymmetric shape of its broad 2.7--3.6~$\mu$m absorption. Its spectrum also shows the strongest resemblance to the spectra of less-red Trojans out of all the small body spectra we have presented. Okyrhoe has one of the closest current orbits among the Centaurs hitherto observed by JWST, and we hypothesize that the difference between Okyrhoe and the more distant Centaur 2003~WL7, which has a spectrum that is much more consistent with that of typical KBOs, is indicative of differential evolution of Centaurs as a function of their received insolation. Within the framework of dynamical instability models of solar system evolution, we can apply this argument to the entire Trojan population, which has been located at 5~au for $\sim$4~Gyr. It follows that the cumulative thermal processing of Trojan surfaces may have further attenuated the 3~$\mu$m OH band beyond the level apparent on Okyrhoe, producing the final spectra we see today. 

It is important to note that the KBOs observed by JWST so far are significantly larger than the Trojans and Centaurs. As such, we are unable to disentangle possible size--dependent trends in the initial bulk composition and subsequent surface evolution of these objects directly. Future JWST observations of additional low-perihelion Centaurs and KBOs smaller than 300~km in diameter will enable a more robust assessment of the proposed evolutionary trajectory that links the KBOs and Centaurs to the Trojans and provide a definitive empirical test of the purported outer solar system origin of Trojans.

\subsection{The 3.2--4.0~$\mu$m region}
\label{subsec:3.4micron}

\begin{figure}
    \centering
    \includegraphics[width=\columnwidth]{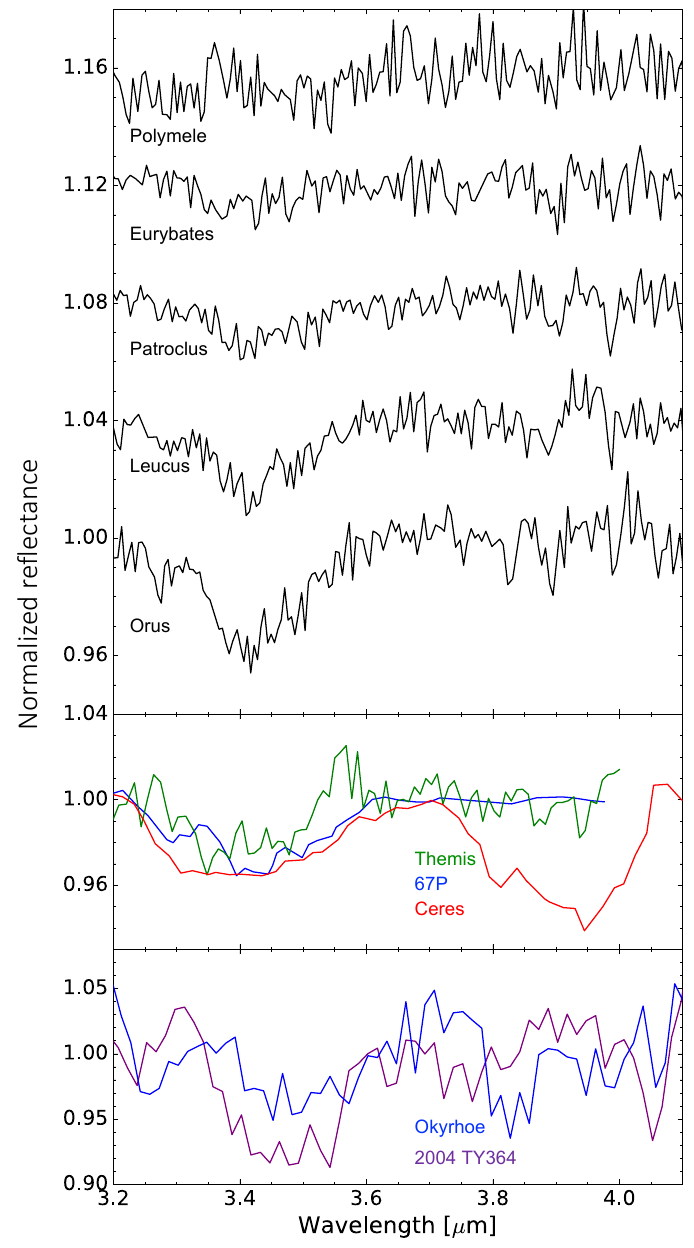}
    \caption{Top panel: a zoomed-in view of the binned Trojan reflectance spectra between 3.2 and 4.1~$\mu$m. The continuum shape has been removed to isolate the 3.3--3.6~$\mu$m aliphatic organic band. The Trojan spectra are arranged in order of increasing band strength. This organic feature is most prominent on the red objects Orus and Leucus. Middle panel: continuum-subtracted spectra of Ceres, 67P, and Themis (binned for clarity), which display a similar 3.4~$\mu$m aliphatic organic feature. 67P and Ceres show evidence for additional aromatic C--H absorption at around 3.3~$\mu$m. The broad 3.8--4.0~$\mu$m absorption in the Ceres spectrum is indicative of carbonate-bearing species; this feature is not detected in any of the Trojan spectra. Bottom panel: binned spectra of Okyrhoe and 2004~TY364, showing aliphatic organic features.}
    \label{fig:3.4micron}
\end{figure}

Moving to slightly longer wavelengths, we find a distinct absorption band centered around 3.4~$\mu$m. In contrast to the aforementioned 3~$\mu$m band, this feature is more prominent on the red objects Orus and Leucus than it is on the less-red targets. To examine the shape of these features better, we fit polynomial continuum functions to the spectra from 3.1 to 4.2~$\mu$m, excluding the region between 3.3 and 3.7~$\mu$m. Figure~\ref{fig:3.4micron} shows the continuum-subtracted spectra, where the 3.4~$\mu$m band is clearly discernible on all targets except for Polymele; it is possible that the relatively poor signal-to-noise ratio of the Polymele spectrum may be obscuring a weak absorption in this wavelength range that has a similar depth to the analogous features seen on the other less-red Trojans. To characterize these bands quantitatively, we first divided the spectra by the local linear continuum fits, anchored to the edges of the band ($\sim$3.25 and 3.57 $\mu$m). We then used a Gaussian to model the band and followed the same 10,000-iteration Monte Carlo procedure as with the 3.0~$\mu$m band. The resulting band center and depth values are reported in Table~\ref{tab:bands}.

Analogous spectral features have been observed across a wide range of solar system small body populations, including main belt asteroids \citep[e.g.,][]{campins2010,rivkin2010,licandro2011,takir2012,desanctis2018}, the Jupiter-family comet 67P \citep{raponi2020}, regular and irregular Jovian and Saturnian satellites \citep[e.g.,][]{mccord1997,cruikshank2008,cruikshank2014,clark2012,brown2014}, and KBOs \citep[e.g.,][]{grundy2016,stern2019,brown2023}. In Figure~\ref{fig:3.4micron}, we plot the continuum-subtracted spectra of Ceres, Themis, 67P, Okyrhoe, and 2004~TY364. All of these spectra show a distinct 3.3--3.6~$\mu$m absorption feature that is characteristic of C--H stretch modes from aliphatic organics. Several individual component bands in this region are resolved in the high-precision Rosetta spectrum of 67P \citep{raponi2020}, including the asymmetric stretching modes of the aliphatic CH$_3$ (3.38~$\mu$m) and CH$_2$ (3.42~$\mu$m) groups and the blended symmetric stretching band of CH$_2$ and CH$_3$ (3.47~$\mu$m). The presence of a distinct absorption feature at 3.3~$\mu$m in the spectrum of 67P and the relative broadening of Ceres' organic absorption band toward shorter wavelengths have been attributed to the presence of aromatic hydrocarbons in addition to the aliphatic organics \citep[e.g.,][]{desanctis2019,raponi2020}. No evidence for aromatic hydrocarbons can be found in the Trojan, KBO, and Centaur spectra. 

Laboratory experiments have demonstrated that aliphatic organic compounds are readily produced through ion bombardment and ultraviolet irradiation of ice mixtures containing various combinations of H$_2$O, CH$_4$, CH$_3$OH, CO, NH$_3$, and N$_2$ \citep[e.g.,][]{strazzulla2001,moore2003,palumbo2004,cruikshank2005,materese2014,urso2020}. An alternative pathway for forming aliphatic organics is through global aqueous alteration, a scenario that has been proposed for even small primitive objects, such as the parent bodies of the carbonaceous chondrite meteorites \citep[e.g.,][]{schulte2004,vinogradoff2018}. In addition to yielding the 3.3--3.6~$\mu$m absorption bands observed throughout the middle and outer solar system, these processes also readily reproduce the reddened and darkened surfaces that are characteristic of Trojans, KBOs, Centaurs, and irregular satellites. 

Within this interpretation, the systematically stronger aliphatic organic bands on Orus and Leucus suggest that the primordial surface composition of red Trojans was significantly richer in CH-bearing species than the less-red Trojans. The complex interplay between relative ice abundance, grain size, and radiation dose that is evident from these past experiments, as well as the multiplicity of possible chemical pathways that produce the same aliphatic organic compounds, makes it impossible to link the observed organic absorption on Trojans to a specific initial ice mixture. While initial surface ice inventories incorporating hypervolatile species such as N$_2$ and CH$_4$ necessitate a cold outer solar system origin, the ability of mixtures containing CH$_3$OH to reproduce the broad 2.7--3.6~$\mu$m spectral feature means that a middle solar system origin for Trojans cannot be fully excluded, as methanol ice can persist on airless surfaces at 5--20~au on kiloyear to megayear timescales \citep[e.g.,][]{wong2016,lisse2022}. 

In the spectrum of Ceres, the 3.4~$\mu$m band is accompanied by an additional major absorption feature at 3.7--4.0~$\mu$m. Past modeling of Ceres' surface has revealed that this pair of bands is indicative of organic compounds mixed with a significant amount of carbonate-bearing species, which absorb strongly at both 3.4 and 3.9--4.0~$\mu$m \citep{desanctis2015}. The longer-wavelength band is particularly prominent within the bright regions in Occator~Crater, where the absorption takes on a distinctive triangular shape with a reflectance minimum around 3.95~$\mu$m \citep{desanctis2016,carrozzo2018}. The intrinsic scatter in the Trojan spectra throughout this wavelength range is poorer than at shorter wavelengths, and there are signs of increased correlated noise features. We do not detect any hints of a broad dip in reflectance and place upper limits of a few percent on the presence of a carbonate absorption at 3.7--4.0~$\mu$m.

\subsection{4.25~$\mu$m CO$_2$ band}
\label{subsec:4.2micron}

\begin{figure}
    \centering
    \includegraphics[width=\columnwidth]{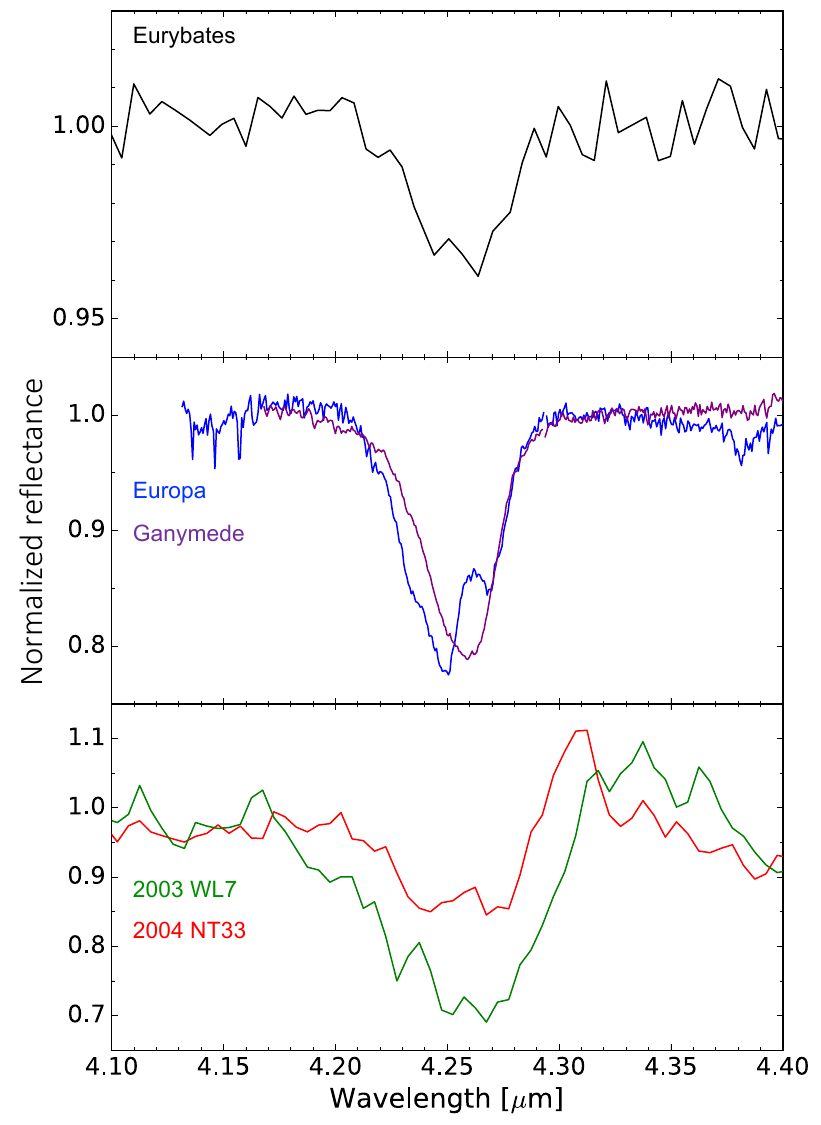}
    \caption{A collection of spectra that show the 4.25~$\mu$m CO$_2$ band. Top panel: binned continuum-subtracted spectra of Eurybates. Middle panel: high spectral resolution JWST spectra of Europa and Ganymede, from \citet{trumbo2023europa} and \citet{trumbo2023}, respectively. Bottom panel: JWST spectra of 2003~WL7 and 2004~NT33.}
    \label{fig:4.2micron}
\end{figure}

The reflectance spectrum of Eurybates contains a distinct absorption band at 4.25~$\mu$m that is not seen on any other Trojan. This wavelength corresponds to the $\nu_3$ vibrational mode of CO$_2$ ice \citep[e.g.,][]{sandford1990,hansen1997}. Analogous spectral features are ubiquitous across the middle and outer solar system, from giant planet satellites \citep[e.g.,][]{mccord1998,cruikshank2010,pinillaalonso2011} to Centaurs and KBOs \citep[e.g.,][]{cruikshank1993,harringtonpinto2022,depra2023,pinillaalonso2023}. Figure~\ref{fig:4.2micron} shows a sample of reflectance spectra displaying 4.25~$\mu$m absorption, plotted alongside the continuum-subtracted spectrum of Eurybates; the high spectral resolution JWST/NIRSpec spectra of Europa and Ganymede were taken from \citet{trumbo2023europa} and \citet{trumbo2023}, respectively. We analyzed the CO$_2$ band shape on Eurybates using a similar methodology to our analysis of the 3.4~$\mu$m band (Section~\ref{subsec:3.4micron}). After subtracting the best-fit linear continuum through the edges of the band at $\sim$4.15 and 4.33~$\mu$m, we carried out a Monte Carlo analysis using a Gaussian band profile. The band center and depth are reported in Table~\ref{tab:bands}.

On the cold surfaces of KBOs and Centaurs, this band, centered around 4.268~$\mu$m, is attributed to pure CO$_2$ ice \citep[e.g.,][]{depra2023}. However, at the typical temperatures of the Jupiter Trojans, CO$_2$ ice is extremely unstable, sublimating on day-long timescales \citep[e.g.,][]{lisse2022}. Any exposed CO$_2$ ice on the surface would be ephemeral, such as in the case of 67P, where isolated patches of CO$_2$ ice in the southern hemisphere were found to persist during its cold season, when the local region was in shadow, before sublimating away on week-long timescales upon exposure to sunlight \citep{filacchione2016}. Another example illustrating the instability of CO$_2$ ice in the Trojan region is the recent JWST observation of the active Centaur 39P/Oterna, which revealed CO$_2$ gas emission while the target was at 5.82~au \citep{39P}. It follows that the CO$_2$ on the surface Eurybates must be bound to some refractory component. Type II CO$_2$ clathrates (CO$_2$ bound in a crystalline water ice structure) are a salient possibility and have been forwarded as an explanation for the 4.27~$\mu$m band seen on Phoebe \citep{cruikshank2010}. Laboratory studies have shown that the CO$_2$ band center for type II clathrates drifts to shorter wavelengths with increasing temperature, reaching $\sim$4.265 $\mu$m at the typical surface temperatures of Trojans \citep{blake1991}. Meanwhile, shorter-wavelength absorption features may be due to mineral-bound CO$_2$ \citep{hibbitts2000,hibbitts2003} or irradiation-formed carbonic acid \citep{jones2014}, which have band centers near 4.25~$\mu$m. Our measured band center for Eurybates ($4.258 \pm 0.002$~$\mu$m) is broadly consistent with both of these possibilities. CO$_2$ can also be produced as a component of the organic-rich residue that forms from the irradiation of carbon-rich ice mixtures --- the same process that yields the 3.4~$\mu$m aliphatic organic features described in Section~\ref{subsec:3.4micron} \citep{urso2020}.

The absence of a discernible CO$_2$ feature on the other Trojan targets --- both less-red and red --- presents an interesting conundrum. Eurybates is the only member of a collisional family among the five targets observed with JWST, so we posit that its visible CO$_2$ absorption is related to its impact history. It is important to note that the estimated age of the Eurybates family (1--4~Gyr; \citealt{marschall2022}) is significantly longer than typical space weathering timescales for Trojans provided in the literature, which are on the order of $10^{2--3}$~yr \citep[e.g.,][]{melita2015}. Therefore, with respect to irradiation-related alteration of surface materials, Eurybates is considered mature. It follows that the differences between Eurybates and the other Trojans must be due to fundamental distinctions in the physical and/or chemical properties of the surfaces.

Here, we once again use the unique spectrum of Okyrhoe to develop a hypothesis for this phenomenon. While Okyrhoe shows a muted 3~$\mu$m feature with a depth that is intermediate between that of Trojans and KBOs, its spectrum lacks a discernible CO$_2$ absorption band. A subset of Centaurs has been observed to display cometary activity, with the activity concentrated among low-perihelion objects \citep[e.g.,][]{jewitt2009,jewitt2015,wong2019centaurs}. The exact trigger for the onset of Centaur activity is still not fully understood, but a leading hypothesis points toward the transition between amorphous to crystalline water ice and the concomitant release of trapped volatiles \citep[e.g.,][]{capria2000,jewitt2009,wierzchos2017}. While Okyrhoe has never been observed to be active, its relatively close orbit places it in a region where the amorphous-to-crystalline transition can take place \citep{guilbertlepoutre2012}. Physical modeling of cometary activity has demonstrated that fine-grained refractory material is entrained in the outgassed volatiles and can be redeposited across the surface \citep[e.g.,][]{jewitt2002}. The obscuration of Okyrhoe's surface due to past activity may explain the distinctly muted 3~$\mu$m OH band and the absence of 4.25~$\mu$m CO$_2$ absorption in its spectrum.

If Trojans initially formed as icy planetesimals in the primordial trans-Neptunian region, as current dynamical instability models suggest, they likely became active during their inward migration from the outer solar system. The increased temperatures and space weathering would have led to thermal processing of the initially volatile-rich surfaces. Due to the observed prevalence of CO$_2$ ice across the KBO population \citep{depra2023,pinillaalonso2023}, it is possible that all Trojans developed surfaces rich in bound CO$_2$ (e.g., through the aforementioned irradiation process). Meanwhile, the activated outgassing at these higher temperatures would have caused simultaneous physical obscuration of the spectral features across the population. In the special case of Eurybates, the later family forming impact destroyed the deposited lag material on the parent body and exposed material from deeper layers, thereby providing an unobscured view of the 4.25~$\mu$m bound CO$_2$ band. Analogous JWST observations of other members of the Eurybates family will enable us to determine whether the CO$_2$ feature is indeed a distinctive feature of the collisional fragments. Recent photometric observations have confirmed a second collisional family within the Trojans --- the Ennomos family \citep{wong2023}. Comparisons of the two families could assess whether there are systematic differences in the way CO$_2$ is bound to the surfaces of Trojans.

\section{Conclusions}
\label{sec:concl}
We have presented new 1.7--5.3~$\mu$m JWST/NIRSpec reflectance spectra of the five Jupiter Trojans that will be visited by the Lucy spacecraft in 2027--2033. These observations provide enhanced sensitivity and wavelength coverage relative to previously published results. Across the five spectra, we have identified and characterized three major absorption bands, which we summarize below.
\begin{enumerate}
    \item A broad OH absorption feature centered around 3~$\mu$m is discernible across the five targets. The feature is systematically deeper on the less-red Trojans, particularly Eurybates, and has a sharp short-wavelength edge at roughly 2.7~$\mu$m. This band is inconsistent with the sharper phyllosilicate feature seen on many large main belt asteroids (e.g., Ceres, Pallas, and Themis), instead being more comparable to the rounded absorption features that are characteristic of water-ice-rich Centaurs and KBOs, as well as the active outer main belt asteroid 238P. While small patches of exposed water ice may account for the presence of this OH band on Trojans, an alternative explanation is that the OH-bearing components within an irradiation residue were formed from the solar processing of volatile ice mixtures rich in carbon and OH-bearing molecules such as methanol.
    \item Between 3.3 and 3.6~$\mu$m, we find a distinct absorption band that is indicative of aliphatic organic compounds. The depth of this feature is stronger on the red Trojans Orus and Leucus than on the less-red object, suggesting that the red Trojans initially formed with a higher relative abundance of CH-bearing species on their surfaces. We do not find evidence for carbonates on the Trojans, which lack the additional absorption band at 3.7--4.0~$\mu$m.
    \item A 4.25~$\mu$m absorption band is present only in the spectrum of Eurybates. This feature indicates the presence of CO$_2$ on the surface; as pure CO$_2$ ice is highly unstable to sublimation loss at 5.2~au, the CO$_2$ on Eurybates is likely bound to or trapped within some less volatile material, e.g., water ice clathrates. The observation that the only target to display this CO$_2$ feature is a collisional fragment leads us to hypothesize that CO$_2$ may be ubiquitous across the Trojan population, with the spectral signature being obscured on the uncollided objects due to some overlaying lag material.
\end{enumerate}

Although these JWST spectra have significantly enriched our understanding of the Trojans' surface composition, we do not yet have definitive proof that exclusively links these objects to the outer solar system, as predicted by recent dynamical instability models. While all of the three spectral features outlined above are commonplace throughout the KBO and Centaur population, the existence of other nearby small bodies with analogous features (e.g., CO$_2$ on Jovian and Saturnian satellites) demonstrates that there may be active chemical processes within the giant planet region that are capable of producing the observed spectral features on Trojans. Nevertheless, the tentative trend between decreasing OH band strength and decreasing heliocentric distance across the KBOs, Centaurs, and Trojans suggests a unified evolutionary trajectory. In particular, the strongly attenuated OH absorption on the close-in Centaur Okyrhoe, with a band depth that is intermediate between those seen on KBOs and Trojans, points toward secondary thermal processing as a potential explanation for the observed variation in spectral shape. Possible activation of outgassing on the Trojans during their purported inward migration, analogous to the process on present day active Centaurs, may also account for the absence of a discernible CO$_2$ band on all but the collisional fragments within the Trojan population.

Looking ahead to the Lucy flybys, we expect that these JWST/NIRSpec spectra will continue to provide crucial insights. For example, because the onboard instruments do not cover the 4--5~$\mu$m region, our knowledge of the unique CO$_2$ band on Eurybates will be indispensable when assessing any systematic differences that are observed at shorter wavelengths in comparison to the other flyby targets. Conversely, the fine spatial resolution provided by Lucy across the Trojans' surfaces will refine our interpretations of the JWST spectra. The spacecraft will be able to probe for isolated patches of exposed water ice and yield compositional maps that link the OH and aliphatic organic bands to specific geological features. Likewise, using thermal infrared measurements, we will unveil the detailed physical properties of the regolith and search for the telltale signs of past cometary activity on Trojans. Ultimately, the findings from the Lucy mission must be synthesized with spectroscopic results for Trojans, Centaurs, KBOs, and other small bodies in order to place the Trojans within the broader picture of solar system formation and evolution properly.

\begin{acknowledgments}
This work is based on observations made with the NASA/ESA/CSA James Webb Space Telescope. The JWST data were obtained from the Mikulski Archive for Space Telescopes (MAST) at the Space Telescope Science Institute, which is operated by the Association of Universities for Research in Astronomy, Inc., under NASA contract NAS 5-03127 for JWST. These observations are associated with program \#2574. The specific observations analyzed can be accessed via \dataset[DOI: 10.17909/hjjb-b241]{https://doi.org/10.17909/hjjb-b241}. The authors thank Michael Kelley for providing the JWST/NIRSpec spectrum of 238P/Read from program \#1252.
\end{acknowledgments}
\facilities{JWST/NIRSpec.}
\software{\texttt{astropy} \citep{astropy2013,astropy2018,astropy2022}, \texttt{jwst} \citep{bushouse2022}, \texttt{matplotlib} \citep{hunter2007}, \texttt{numpy} \citep{harris2020}, and \texttt{scipy} \citep{virtanen2020}}


{}

\begin{thebibliography}{}
\tighten
\footnotesize

\bibitem[Astropy Collaboration(2022)]{astropy2022}
Astropy Collaboration, Price-Whelan, A.~M., Lim, P.~L., et~al. 2022, \apj, 935, 167

\bibitem[Astropy Collaboration(2013)]{astropy2013}
Astropy Collaboration, Price-Whelan, A.~M., Sip{\H o}cz, B.~M., et~al. 2018, \aj, 156, 123 

\bibitem[Astropy Collaboration(2018)]{astropy2018}
Astropy Collaboration, Robitaille, T.~P., Tollerud, E.~J., et~al. 2013, \aap, 558, A33

\bibitem[Baratta \& Palumbo(1998)]{baratta1998}
Baratta, G., \& Palumbo, M. 1998, JOSAA, 15, 3076

\bibitem[Barkume et~al.(2008)]{barkume2008}
Barkume, K.~M., Brown, M.~E., \& Schaller, E.~L. 2008, \aj, 135, 55

\bibitem[Barucci et~al.(2011)]{barucci2011}
Barucci, M.~A., Alvarez-Candal, A., Merlin, F., et~al. 2011, Icar, 214, 297

\bibitem[Barucci et~al.(2016)]{barucci2016}
Barucci, M.~A., Filacchione, G., Fornasier, S., et~al. 2016, \aap, 595, A102

\bibitem[Blake et~al.(1991)]{blake1991}
Blake, D., Allamandola, L., Sandford, S., Hudgins, D., \& Freund, F. 1991, Sci, 254, 548

\bibitem[B{\"o}ker et~al.(2023)]{boker2023}
B{\"o}ker, T., Beck, T.~L., Birkmann, S.~M., et~al. 2023, PASP, 135, 038001

\bibitem[Brown(2016)]{brown2016}
Brown, M.~E. 2016, \aj, 152, 159

\bibitem[Brown \& Fraser(2023)]{brown2023}
Brown, M.~E., \& Fraser, W.~C. 2023, PSJ, 4, 130

\bibitem[Brown \& Rhoden(2014)]{brown2014}
Brown, M.~E., \& Rhoden, A.~R. 2014, \apjl, 793, L44

\bibitem[Bro{\v z} \& Rozehnal(2011)]{broz2011}
Bro{\v z}, M., \& Rozehnal, J. 2011, \mnras, 414, 565

\bibitem[Buie et~al.(2021)]{buie2021}
Buie, M.~W., Keeney, B.~A., Strauss, R.~H., et~al. 2021, PSJ, 2, 202

\bibitem[Buie et~al.(2018)]{buie2018}
Buie, M.~W., Zangari, A.~M., Marchi, S., Levison, H.~F., \& Mottola, S. 2018, \aj, 155, 245

\bibitem[Buratti et~al.(2008)]{buratti2008}
Buratti, B.~J., Soderlund, K., Bauer, J., et~al. 2008, Icar, 193, 309

\bibitem[Bushouse et~al.(2022)]{bushouse2022}
Bushouse, H., Eisenhamer, J., Dencheva, N., et~al. 2023, JWST Calibration Pipeline, v1.11.3, Zenodo, doi:10.5281/zenodo.8157276

\bibitem[Campins et~al.(2010)]{campins2010}
Campins, H., Hargrove, K., Pinilla-Alonso, N., et~al. 2010, Natur, 464, 1320

\bibitem[Capria et~al.(2000)]{capria2000}
Capria, M.~T., Coradini, A., De Sanctis, M.~C., \& Orosei, R. 2000, \aj, 119, 3112

\bibitem[Carrozzo et~al.(2018)]{carrozzo2018}
Carrozzo, F.~G., De Sanctis, M.~C., Raponi, A., et~al. 2018, SciA, 4, e1701645

\bibitem[Clark et~al.(2005)]{clark2005}
Clark, R.~N., Brown, R.~H., Jaumann, R. et~al. 2005, Natur, 435, 66

\bibitem[Clark et~al.(2012)]{clark2012}
Clark, R.~N., Cruikshank, D.~P., Jaumann, R., et~al. 2012, Icar, 218, 831

\bibitem[Cruikshank et~al.(2014)]{cruikshank2014}
Cruikshank, D.~P., Dalle Ore, C.~M., Clark, R.~N., \& Pendelton, Y.~J. 2014, Icar, 233, 306

\bibitem[Cruikshank et~al.(2005)]{cruikshank2005}
Cruikshank, D.~P., Imanaka, H., \& Dalle Ore, C.~M. 2005, AdSpR, 36, 178

\bibitem[Cruikshank et~al.(2010)]{cruikshank2010}
Cruikshank, D.~P., Meyer, A.~W., Brown, R.~H., et~al. 2010, Icar, 206, 561

\bibitem[Cruikshank et~al.(1993)]{cruikshank1993}
Cruikshank, D.~P., Roush, T.~L., Owen, T.~C., et~al. 1993, Sci, 261, 742

\bibitem[Cruikshank et~al.(2008)]{cruikshank2008}
Cruikshank, D.~P., Wegryn, E., Dalle Ore, C.~M., et~al. 2008, Icar, 193, 334

\bibitem[de Pr{\'a} et~al.(2024)]{depra2023}
de Pr{\'a}, M.~N., H{\'e}nault, E., Pinilla-Alonso, N., et~al. 2024, NatAs, in press

\bibitem[De Sanctis et~al.(2018)]{desanctis2018}
De Sanctis, M.~C., Ammannito, E., Carrozzo, F.~G., et~al. 2018, M\&PS, 53, 1844

\bibitem[De Sanctis et~al.(2015)]{desanctis2015}
De Sanctis, M.~C., Ammannito, E., Raponi, A., et~al. 2015, Natur, 528, 241

\bibitem[De Sanctis et~al.(2016)]{desanctis2016}
De Sanctis, M.~C., Raponi, A., Ammannito, E., et~al. 2016, Natur, 536, 54

\bibitem[De Sanctis et~al.(2019)]{desanctis2019}
De Sanctis, M.~C., Vinogradoff, V., Raponi, A., et~al. 2019, \mnras, 482, 2407

\bibitem[Dotto et~al.(2006)]{dotto2006}
Dotto, E., Fornasier, S., Barucci, M.~A., et~al. 2006, Icar, 183, 420

\bibitem[Duffard et~al.(2014)]{duffard2014}
Duffard, R., Pinilla-Alonso, N., Santos-Sanz, P., et~al. 2014, \aap, 564, A92

\bibitem[Emery \& Brown(2003)]{emery2003}
Emery, J.~P., \& Brown, R.~H. 2003, Icar, 164, 104

\bibitem[Emery et~al.(2011)]{emery2011}
Emery, J.~P., Burr, D.~M., \& Cruikshank, D.~P. 2011, \aj, 141, 25

\bibitem[Emery et~al.(2024)]{emery2023}
Emery, J.~P., Wong, I., Brunetto, R., et~al. 2024, Icar, in press (arXiv:2309.15230)

\bibitem[Filacchione et~al.(2016)]{filacchione2016}
Filacchione, G., Raponi, A., Capaccioni, R., et~al. 2016, Sci, 354, 1563


\bibitem[Fornasier et~al.(2007)]{fornasier2007}
Fornasier, S., Dotto, E., Hainaut, O., et~al. 2007, Icar, 190, 622

\bibitem[Fraser \& Brown(2012)]{fraser2012}
Fraser, W.~C., \& Brown, M.~E. 2012, \apj, 749, 33

\bibitem[Fraser \& Brown(2018)]{fraser2018}
Fraser, W.~C., \& Brown, M.~E. 2018, \aj, 156, 23

\bibitem[Fraser et~al.(2023)]{fraser2023}
Fraser, W.~C., Pike, R.~E., Marsset, M., et~al. 2023, PSJ, 4, 80

\bibitem[Gomes et~al.(2005)]{gomes2005}
Gomes, R., Levison, H.~F., Tsiganis, K., \& Morbidelli, A. 2005, Natur, 435, 466


\bibitem[Grundy et~al.(2016)]{grundy2016}
Grundy, W.~M., Binzel, R.~P., Buratti, B.~J., et~al. 2016, Sci, 351, 9189

\bibitem[Grundy et~al.(2018)]{grundy2018}
Grundy, W.~M., Noll, K.~S., Buie, M.~W., \& Levison, H.~F. 2018, Icar, 305, 198

\bibitem[Grundy et~al.(2024)]{grundy2023}
Grundy, W.~M., Wong, I., Glein, C.~R., et~al. 2024, Icar, 411, 115923

\bibitem[Guilbert-Lepoutre(2012)]{guilbertlepoutre2012}
Guilbert-Lepoutre, A. 2012, \aj, 144, 97

\bibitem[Guilbert-Lepoutre(2014)]{guilbertlepoutre2014}
Guilbert-Lepoutre, A. 2014, Icar, 231, 232

\bibitem[Haghighipour(2009)]{haghighipour2009}
Haghighipour, N. 2009, M\&PS, 44, 1863

\bibitem[Hansen(1997)]{hansen1997}
Hansen, G.~B. 1997, JGR, 102, 21569

\bibitem[Harrington Pinto et~al.(2023)]{39P}
Harrington Pinto, O., Kelley, M.~S.~P., Villanueva, G.~L., et~al. 2023, PSJ, 4, 208

\bibitem[Harrington Pinto et~al.(2022)]{harringtonpinto2022}
Harrington Pinto, O., Womack, M., Fernandez, Y., \& Bauer, J. 2022, PSJ, 3, 247

\bibitem[Harris(1998)]{harris1998}
Harris, A.~W. 1998, Icar, 131, 291

\bibitem[Harris et~al.(2020)]{harris2020}
Harris, C.~R., Millman, K.~J., van der Walt, S.~J., et~al. 2020, Natur, 585, 357

\bibitem[Hibbitts et~al.(2000)]{hibbitts2000}
Hibbitts, C.~A., McCord, T.~B., \& Hansen, G.~B. 2000, JGR, 105, 22541

\bibitem[Hibbitts et~al.(2003)]{hibbitts2003}
Hibbitts, C.~A., Pappalardo, R.~T., Hansen, G.~B., \& McCord, T.~B. 2003 JGRE, 108, 5036

\bibitem[Hsieh et~al.(2016)]{hsieh2016}
Hsieh, H.~H., \& Haghighipour, N. 2016, Icar, 277, 19

\bibitem[Hunter(2007)]{hunter2007}
Hunter, J.~D. 2007, CSE, 9, 90

\bibitem[Jakobsen et~al.(2022)]{jakobsen2022}
Jakobsen, P., Ferruit, P., Alves de Oliveria, C., et~al. 2022, \aap, 661, A80

\bibitem[Jewitt(2009)]{jewitt2009}
Jewitt, D. 2009, \aj, 137, 4296

\bibitem[Jewitt(2015)]{jewitt2015}
Jewitt, D. 2015, \aj, 150, 201

\bibitem[Jewitt(2002)]{jewitt2002}
Jewitt, D.,~C., 2002, \aj, 123, 1039

\bibitem[Jones et~al.(2014)]{jones2014}
Jones, B.~M., Kaiser, R.~I., \& Strazzulla, G. 2014, \apj, 788, 170

\bibitem[Jones et~al.(1990)]{jones1990}
Jones, T.~D., Lebofsky, L.~A., Lewis, J.~S., \& Marley, M.~S. 1990, Icar, 88, 172

\bibitem[Kelley et~al.(2023)]{kelley2023}
Kelley, M.~S.~P., Hsieh, H.~H., Bodewits, D., et~al. 2023, Natur, 619, 720

\bibitem[King et~al.(1992)]{king1992}
King, T.~V.~V., Clark, R.~N., Calvin, W.~M., Sherman, D.~M., \& Brown, R.~H. 1992, Sci, 255, 1551

\bibitem[Lebofsky(1980)]{lebofsky1980}
Lebofsky, L.~A. 1980, \aj, 85, 573

\bibitem[Lellouch et~al.(2013)]{lellouch2013}
Lellouch, E., Santos-Sanz, P., Lacerda, P., et~al. 2013, \aap, 557, A60

\bibitem[Levison et~al.(2008)]{levison2008}
Levison, H.~F., Morbidelli, A., Van Laerhoven, C., Gomes, R., \& Tsiganis, K. 2008, Icar, 196, 258

\bibitem[Levison et~al.(2021)]{levison2021}
Levison, H.~F., Olkin, C.~B., Noll, K.~S., et~al. 2021, PSJ, 2, 171

\bibitem[Licandro et~al.(2011)]{licandro2011}
Licandro, J., Campins, H., Kelley, M., et~al. 2011, \aap, 525, A34

\bibitem[Licandro et~al.(2024)]{licandro2023}
Licandro, J., Pinilla-Alonso, N., Holler, B.~J., et~al. 2024, NatAs, in revision


\bibitem[Lisse et~al.(2022)]{lisse2022}
Lisse, C.~M., Gladstone, G.~R., Young, L.~A., et~al. 2022, PSJ, 3, 112

\bibitem[Marchi et~al.(2023)]{marchi2023}
Marchi, S., Nesvorn{\'y}, D., Vokrouhlick{\'y}, D., Bottke, W.~F., \& Levison, H.~F. 2023, \aj, 166, 221

\bibitem[Marschall et~al.(2022)]{marschall2022}
Marschall, R., Nesvorn{\'y}, D., Deienno, R., et~al. 2022, \aj, 164, 167

\bibitem[Marzari \& Scholl(1998)]{marzari1998}
Marzari, F., \& Scholl, H. 1998, \aap, 339, 278

\bibitem[Marzari et~al.(2003)]{marzari2003}
Marzari, F., Tricarico, P., \& Scholl, H. 2003, Icar, 162, 453

\bibitem[Mastrapa et~al.(2009)]{mastrapa2009}
Mastrapa, R.~M., Sandford, S.~A., Roush, T.~L., Cruikshank, D.~P., \& Dalle Ore, C.~M. 2009, \apj, 701, 1347

\bibitem[Mastrapa et~al.(2013)]{mastrapa2013}
Mastrapa, R.~M.~E., Grundy, W.~M., \& Gudipati, M.~S. 2013, in The Science of Solar System Ices, ed. M.~S.~Gudipati \& J.~Castillo-Rogez (Berlin: Springer), 371

\bibitem[Materese et~al.(2014)]{materese2014}
Materese, C.~K., Cruikshank, D.~P., Sandford, S.~A., et~al. 2014, \apj, 788, 111

\bibitem[McCord et~al.(1997)]{mccord1997}
McCord, T.~B., Carlson, R., Smythe, W., et~al. 1997, Sci, 278, 271

\bibitem[McCord et~al.(1998)]{mccord1998}
McCord, T.~B., Hansen, G.~B., Clark, R.~N., et~al. 1998, JGR, 103, 8603

\bibitem[McGraw et~al.(2022)]{mcgraw2022}
McGraw, L.~E., Emery, J.~P., Thomas, C.~A., et~al. 2022, PSJ, 3, 243

\bibitem[McKay et~al.(2017)]{mckay2017}
McKay, A.~J., Bodewits, D., \& Li, J.-Y. 2017, Icar, 286, 308

\bibitem[Melita et~al.(2015)]{melita2015}
Melita, M.~D., Ka{\v n}uchov{\'a}, Z., Brunetto, R., \& Strazzulla, G. 2015, Icar, 248, 222

\bibitem[Moore \& Hudson(2003)]{moore2003}
Moore, M.~H., \& Hudson, R.~L. 2003, Icar, 161, 486

\bibitem[Morbidelli et~al.(2005)]{morbidelli2005}
Morbidelli, A., Levison, H.~F., Tsiganis, K., \& Gomes, R. 2005, Natur, 435, 462

\bibitem[Moseley et~al.(2010)]{moseley2010}
Moseley, S., Arendt, R.~G., Fixsen, D.~J., et~al. 2010, SPIE, 7742, 77421B

\bibitem[Mottola et~al.(2023)]{mottola2023}
Mottola, S., Hellmich, S., Buie, M.~W., et~al. 2023, PSJ, 4, 18

\bibitem[Mueller et~al.(2010)]{mueller2010}
Mueller, M., Marchis, F., Emery, J.~P., et~al. 2010, Icar, 205, 505

\bibitem[Nesvorn{\'y} et~al.(2013)]{nesvorny2013}
Nesvorn{\'y}, D., Vokrouhlick{]'y}, D., \& Morbidelli, A. 2013, \apj, 768, 45


\bibitem[Olkin et~al.(2021)]{olkin2021}
Olkin, C.~B., Levison, H.~F., Vincent, M., et~al. 2021, PSJ, 2, 172

\bibitem[O'Rourke et~al.(2020)]{orourke2020}
O'Rourke, L., M{\"u}ller, T.~G., Biver, N., et~al. 2020, \apjl, 898, L45

\bibitem[Palumbo et~al.(2004)]{palumbo2004}
Palumbo, M.~E., Ferini, G., \& Baratta, G.~A. 2004, AdSpR, 33, 49

\bibitem[Perrin et~al.(2014)]{perrin2014}
Perrin, M.~D., Sivaramakrishnan, A., Lajoie, C.-P., et~al. 2014, Proc. SPIE, 9143, 91433X

\bibitem[Pinilla-Alonso et~al.(2024)]{pinillaalonso2023}
Pinilla-Alonso, N., Brunetto, R., de Pr{\'a}, M.~N., et~al. 2024, NatAs, in revision

\bibitem[Pinilla-Alonso et~al.(2011)]{pinillaalonso2011}
Pinilla-Alonso, N., Roush, T.~L., Marzo, G.~A., et~al. 2011, Icar, 215, 75

\bibitem[Poch et~al.(2020)]{poch2020}
Poch, O., Istiqomah, I., Quirico, E., et~al. 2020, Sci, 367, 1212

\bibitem[Protopapa et~al.(2014)]{protopapa2014}
Protopapa, S., Sunshine, J.~M., Feaga, L.~M., et~al. 2014, Icar, 238, 191

\bibitem[Raponi et~al.(2020)]{raponi2020}
Raponi, A., Ciarniello, M., Capaccioni, F., et~al. 2020, NatAs, 4, 500

\bibitem[Quirico \& Schmitt(1997)]{quirico1997}
Quirico, E., \& Schmitt, B. 1997, Icar, 127, 354

\bibitem[Rauscher et~al.(2012)]{rauscher2012}
Rauscher, B.~J., Arendt, R.~G., Fixsen, D.~J., et~al. 2012, Proc. SPIE, 8453, 84531F

\bibitem[Rivkin \& Emery(2010)]{rivkin2010}
Rivkin, A.~S., \& Emery, J.~P. 2010, Natur, 464, 1322

\bibitem[Rivkin et~al.(2002)]{rivkin2002}
Rivkin, A.~S., Howell, E.~S., Vilas, F., \& Lebofsky, L.~A. 2002, in Asteroids III, ed. W.~F. Bottke Jr., et al. (Univ. of Arizona), 235

\bibitem[Roig \& Nesvorn{\'y}(2015)]{roig2015}
 Roig, F., \& Nesvorn{\'y}, D. 2015, \aj, 150 186

\bibitem[Roig et~al.(2008)]{roig2008}
Roig, F., Ribeiro, A.~O., \& Gil-Hutton, R. 2008, \aap, 483, 911

\bibitem[Sandford \& Allamandola(1990)]{sandford1990}
Sandford, S.~A., \& Allamandola, L.~J. 1990, \apj, 355, 357

\bibitem[Schulte \& Shock(2004)]{schulte2004}
Schulte, M., \& Shock, E. 2004, M\&PS, 39, 1577

\bibitem[Sharkey et~al.(2019)]{sharkey2019}
Sharkey, B.~N.~L., Reddy, V., Sanchez, J.~A., Izawa, M.~R.~M., \& Emery, J.~P. 2019, \aj, 158, 204

\bibitem[Stern et~al.(2019)]{stern2019}
Stern, S.~A., Weaver, H.~A., Spencer, J.~R., et~al. 2019, Sci, 364, 9771

\bibitem[Strazzulla et~al.(2001)]{strazzulla2001}
Strazzulla, G., Baratta, G.~A., \& Palumbo, M.~E. 2001, AcSpA, 57, 825

\bibitem[Sunshine et~al.(2006)]{sunshine2006}
Sunshine, J.~M., A'Hearn, M.~F., Groussin, O., et~al. 2006, Sci, 311, 1453

\bibitem[Szab{\'o} et~al.(2007)]{szabo2007}
Szab{\'o}, Gy. M., Ivezi{\'c}, {\v Z}., Juri{\'c}, M., \& Lupton, R. 2007, \mnras, 377, 1393

\bibitem[Takir \& Emery(2012)]{takir2012}
Takir, D., \& Emery, J.~P. 2012, Icar, 219, 641

\bibitem[Trumbo \& Brown(2023)]{trumbo2023europa}
Trumbo, S.~K., \& Brown, M.~E. 2023, Sci, 381, 1308

\bibitem[Trumbo et~al.(2023)]{trumbo2023}
Trumbo, S.~K., Brown, M.~E., Bockel{\'e}e-Morvan, D., et~al. 2023, SciA, 9, 3724


\bibitem[Tsiganis et~al.(2005)]{tsiganis2005}
Tsiganis, K., Gomes, R., Morbidelli, A., \& Levison, H.~F. 2005, Natur, 435, 459

\bibitem[Urso et~al.(2020)]{urso2020}
Urso, R.~G., Vuitton, V., Danger, G., et~al. 2020, \aap, 644, A115

\bibitem[Usui et~al.(2019)]{usui2019}
Usui, F., Hasegawa, S., Ootsubo, T., \& Onaka, T. 2019, PASJ, 71, 1

\bibitem[Vilenius et~al.(2014)]{vilenius2014}
Vilenius, E., Kiss, C., M{\"u}ller, T., et~al. 2014, \aap, 564, A35

\bibitem[Vinogradoff et~al.(2018)]{vinogradoff2018}
Vinogradoff, V., Bernard, S., Le Guillou, C., \& Remusat, L. 2018, Icar, 305, 358

\bibitem[Virtanen et~al.(2020)]{virtanen2020}
Virtanen, P., Gommers, R., Oliphant, T.~E., et~al. 2020, NatMe, 17, 261

\bibitem[Vokrouhlick{\'y} et~al.(2016)]{vokrouhlicky2016}
Vokrouhlick{\'y}, D., Bottke, W.~F., \& Nesvorn{\'y}, D. 2016, \aj, 152, 39

\bibitem[Wierzchos et~al.(2017)]{wierzchos2017}
Wierzchos, K., Womack, M., \& Sarid, G. 2017, \aj, 153, 230

\bibitem[Wong \& Brown(2015)]{wong2015}
Wong, I., \& Brown, M.~E. 2015, \aj, 150, 174

\bibitem[Wong \& Brown(2016)]{wong2016}
Wong, I., \& Brown, M.~E. 2016, \aj, 152, 90

\bibitem[Wong \& Brown(2017)]{wong2017}
Wong, I., \& Brown, M.~E. 2017, \aj, 153, 145

\bibitem[Wong \& Brown(2019)]{wong2019patroclus}
Wong, I., \& Brown, M.~E. 2019, \aj, 157, 203

\bibitem[Wong et~al.(2019a)]{wong2019uv}
Wong, I., Brown, M.~E., Blacksberg, J., et~al. 2019, \aj, 157, 161

\bibitem[Wong et~al.(2019b)]{wong2019centaurs}
Wong, I., Mishra, A., \& Brown, M.~E., 2019, \aj, 157, 225

\bibitem[Wong \& Brown(2023)]{wong2023}
Wong, I., \& Brown, M.~E. 2023, \aj, 165, 15

\bibitem[Wong et~al.(2014)]{wong2014}
Wong, I., Brown, M.~E., \& Emery, J.~P. 2014, \aj, 148, 112

\bibitem[Yang \& Jewitt(2007)]{yang2007}
Yang, B., \& Jewitt, D. 2007, \aj, 134, 223

\end{thebibliography}
\end{document}